\documentclass[12pt,reqno]{amsart}

\usepackage[active]{srcltx}

\usepackage{latexsym}
\usepackage{graphicx}
\usepackage{subcaption}
\usepackage{amsmath}
\usepackage{mathabx}
\usepackage{hyperref} 
\hypersetup{
    colorlinks,
    citecolor=black,
    filecolor=black,
    linkcolor=black,
    urlcolor=black
}

\newcommand{\Sch}{Schr\"odinger }
\newcommand{\set}[1]{\left\{#1\right\}}
\usepackage{color}

\newcommand{\baa}{\begin{align*}}
\newcommand{\eaa}{\end{align*}}
\newcommand{\bea}{\begin{eqnarray*} }
\newcommand{\eea}{\end{eqnarray*} }
\newcommand{\beq}{\begin{equation} }
\newcommand{\eeq}{\end{equation} }
\newcommand{\bp}{\begin{proposition}}
\newcommand{\ep}{\end{proposition}}
\newcommand{\bt}{\begin{theorem}}
\newcommand{\et}{\end{theorem}}
\newcommand{\bpf}{\begin{proof}}
\newcommand{\epf}{\end{proof}}
\newcommand{\bl}{\begin{lemma}}
\newcommand{\el}{\end{lemma}}
\newcommand{\bc}{\begin{cor}}
\newcommand{\ec}{\end{cor}}
\newcommand{\bd}{\begin{definition}}
\newcommand{\ed}{\end{definition}}

\newcommand{\club}{\clubsuit} 
\newcommand{\cs}{$\club$}

\newcommand{\ihbar}{{\frac{i}{h}}}

\newcommand{\be}{\begin{equation} }
\newcommand{\ee}{\end{equation} }
\newcommand{\bee}{\begin{eqnarray} }
\newcommand{\eee}{\end{eqnarray} }

\newcommand{\gives}{\ensuremath{\rightarrow}}

\newcommand{\x}{\ensuremath{\times}}

\newcommand{\EE}[1]{\mathbb E}
\newcommand{\abs}[1]{\left\lvert #1 \right\rvert}
\newcommand{\norm}[1]{\left\lVert#1\right\rVert}
\newcommand{\lr}[1]{\ensuremath{\left(#1\right)}}

\renewcommand{\Im}{\ensuremath{\mathrm{Im} \ }}

\renewcommand{\Im}{{\operatorname{Im}\,}}

\renewcommand{\epsilon}{\varepsilon}

\newcommand{\wt}{\widetilde}

\newcommand{\R}{{\mathbb R}}
\newcommand{\C}{{\mathbb C}}

\newcommand{\Z}{{\mathbb Z}}

\newcommand{\half}{{\textstyle \frac 12}}

\newcommand{\supp}{{\operatorname{Supp\,}}}
\renewcommand{\phi}{\varphi}

\newcommand{\dcal}{\mathcal{D}}

\newcommand{\fcal}{\mathcal{F}}

\newcommand{\pcal}{\mathcal{P}}

\newcommand{\scal}{\mathcal{S}}

\newcommand{\ucal}{\mathcal{U}}

\newcommand{\wcal}{\mathcal{W}}

\DeclareMathOperator{\Ai}{Ai}

\newtheorem{theo}{{\sc Theorem}}[section]
\newtheorem{lem}[theo]{{\sc Lemma}}
\newtheorem{prop}[theo]{{\sc Proposition}}
\newtheorem{defn}[theo]{{\sc Definition}}

\newtheorem{cor}[theo]{{\sc Corollary}}

\newtheorem{Lem}[theo]{{\sc Lemma}}

\newtheorem{proposition}[theo]{{\sc Proposition}}
\newtheorem{lemma}[theo]{{\sc Lemma}}

\newenvironment{rem}{\medskip\noindent{\it Remark:\/} }{\medskip}

\title[Interface asymptotics of Wigner distributions]
{Interface asymptotics of Wigner-Weyl distributions for the Harmonic Oscillator}

\begin{document}
\author{Boris Hanin and Steve Zelditch}
\address{Department of Mathematics, Texas A\&M and Facebook AI Research, NYC}

\email[B. Hanin]{bhanin@math.tamu.edu}

\address{Department of Mathematics, Northwestern University, Evanston, IL
60208, USA}
\email[S. Zelditch]{zelditch@math.northwestern.edu}

\thanks{Research partially supported by NSF grant DMS-1810747.}
\maketitle

\begin{abstract}  We prove several types of scaling results for Wigner distributions of spectral projections of the isotropic Harmonic oscillator on $\R^d$. In prior work, we studied Wigner distributions $W_{\hbar, E_N(\hbar)}(x, \xi)$ of individual eigenspace projections. In this continuation, we study Weyl sums of such Wigner distributions as the eigenvalue $E_N(\hbar)$ ranges over spectral intervals $[E - \delta(\hbar), E + \delta(\hbar)]$ of various widths $\delta(\hbar)$ and as $(x, \xi) \in T^*\R^d$ ranges over tubes of various widths around the classical energy surface $\Sigma_E \subset T^*\R^d$.  The main results pertain to  interface Airy scaling asymptotics around  $\Sigma_E$, which  divides phase space into an allowed and a forbidden region. The first result pertains
to  $\delta(\hbar) = \hbar$ widths and generalizes our earlier results on Wigner distributions of individual eigenspace projections. Our second result
pertains to $\delta(\hbar) = \hbar^{2/3}$ spectral widths  and Airy asymptotics  of the Wigner distributions in $\hbar^{2/3}$-tubes around  $\Sigma_E$. Our third result pertains to bulk spectral intervals
of fixed width and the behavior of the Wigner distributions inside the energy
surface, outside the energy surface and in a thin neighborhood of the energy surface.

\end{abstract}

\date{\today}

\section{\label{INTRO} Introduction}
\noindent This article is part of a series \cite{HZ19} studying the scaling asymptotics of Wigner distributions 
\begin{equation*}
\sum_{N \in {\mathbb N}} f_{\hbar, E} (E_N(\hbar)) W_{\hbar, E_N(\hbar)}(x, \xi),\;\; (x, \xi) \in T^*\R^d \end{equation*}
of various spectral projection kernels  for the isotropic Harmonic Oscillator,
 \begin{equation} \label{Hh}
\widehat{H}_{h} =  \sum_{j = 1}^d \left(- \frac{\hbar^2}{2}   \frac{\partial^2 }{\partial
x_j^2} + \frac{ x_j^2}{2} \right) \quad\mathrm{on}\quad L^2(\R^d,dx).
\end{equation}
As is well-known,  the spectrum of $\widehat{H}_{\hbar}$ consists of the eigenvalues,
\begin{equation} \label{ENh} E_N(\hbar)=\hbar\lr{N+d/2},\qquad N = 0, 1, 2, \dots.\end{equation} 
We denote the  corresponding eigenspaces by
\begin{equation} \label{VhE} V_{\hbar, E_N(\hbar)}: = \{\psi \in L^2(\R^d): \hat{H}_{\hbar} \psi = 
E_N(\hbar) \psi \},\;\; (\dim V_{\hbar, E_N(\hbar)} \simeq N^{d-1}), \end{equation}
and the eigenspace projections by
\begin{equation} \label{PiDEF} \Pi_{\hbar, E_N(\hbar)}: L^2(\R^d) \to V_{\hbar,E_N(\hbar)}. \end{equation}
The semi-classical Wigner distributions of the projections \eqref{PiDEF} are defined by
\begin{equation}\label{WIGNERDEF1}
 W_{\hbar, E_N(\hbar)}(x, \xi) := \int_{\R^d} \Pi_{\hbar, E_N(\hbar)} \left( x+\frac{v}{2}, x-\frac{v}{2} \right) e^{-\frac{i}{\hbar} v \cdot \xi} \frac{dv}{(2\pi h)^d} .
\end{equation} 
When $E_N(\hbar) = E + o(1)$ as $\hbar \to 0$, $W_{\hbar, E_N(\hbar)}$ is thought of as the `quantization' of the energy surface,
\begin{equation} \label{SIGMAEDEF} \Sigma_E  =\{(x, \xi) \in T^*\R^d: H(x, \xi): = \half(||x||^2 + ||\xi||^2) = E\},  \end{equation}
and \eqref{WIGNERDEF1} is thought of as an approximate $\delta$-function on \eqref{SIGMAEDEF}. This is true in the weak* sense, but the pointwise behavior is quite a bit more complicated and is studied in \cite{HZ19}. The purpose of this article is to study `Weyl sums' of Wigner distributions (the `Wigner-Weyl distributions' of the title)  of the form
\begin{equation} \label{Whfdelta} 
W_{\hbar, f, \delta(\hbar)}(x, \xi): =  \sum_{N} f\lr{\delta(\hbar)^{-1}(E- E_N(\hbar))}W_{ \hbar, E_N}(x, \xi), 
\end{equation} 
where $\delta(h)$ controls the width and $f$ controls the smoothness of the localization. Thus, the essential point is to simultaneously study two types
of `localization': 
\bigskip
\begin{itemize}

\item Spectral localization, determined by the choice of scale $\delta(\hbar)$. For instance, if $\delta(\hbar)=\hbar^\gamma$ and $f$ is the indicator function of the interval $[-a,b]$, then $W_{\hbar, f, \delta(\hbar)}(x, \xi)$ the is the sum of eigenspace Wigner distributions $W_{\hbar, E_N(\hbar)}$ for energies $E_N(\hbar)$ in the spectral interval $[E-b\hbar^\gamma, E+a\hbar^\gamma]$; \bigskip

\item Phase space localization, determined by a choice of region $U \subset T^* \R^d$ over which $(x,\xi)$ can vary. For example, in different results we will consider $U$ to be the $\hbar^{2/3}$ tube around the energy surface $\Sigma_E$ as well as regions deeper in the interior or exterior of $\{H \leq E\}$.  \end{itemize}
\bigskip

In \cite{HZ19} we studied the extreme case of spectral localization, where the window contained just one distinct eigenvalue $E_N(\hbar)$ and it was  assumed that $E_N(\hbar) = E$ for a fixed  $E \in \R_+$.   In particular, we studied the asymptotics of $W_{\hbar, E_N(\hbar)}(x,\xi)$ in different regions around $\Sigma_E$. Roughly, we found with $H(x,\xi)=(\norm{x}^2+\norm{\xi}^2)/2$ denoting the classical Hamiltonian that for fixed $E$,
\[ (2\pi)^d W_{\hbar, E}(x,\xi) ~\simeq~
\begin{cases}
  \hbar^{-d+1/2}H(x,\xi)^{-d/2}\x\mathrm{~oscillatory},&\quad 0<H(x,\xi)< E\\
\hbar^{-d+1/3}\mathrm{Ai}\lr{\hbar^{-2/3}\lr{H(x,\xi)-E}},&\quad H(x,\xi)-E\approx \hbar^{2/3}\\
O(\hbar^\infty),&\quad H(x,\xi)> E
\end{cases}.
\]

\begin{figure}
    \centering
     \includegraphics[width=.5\textwidth]{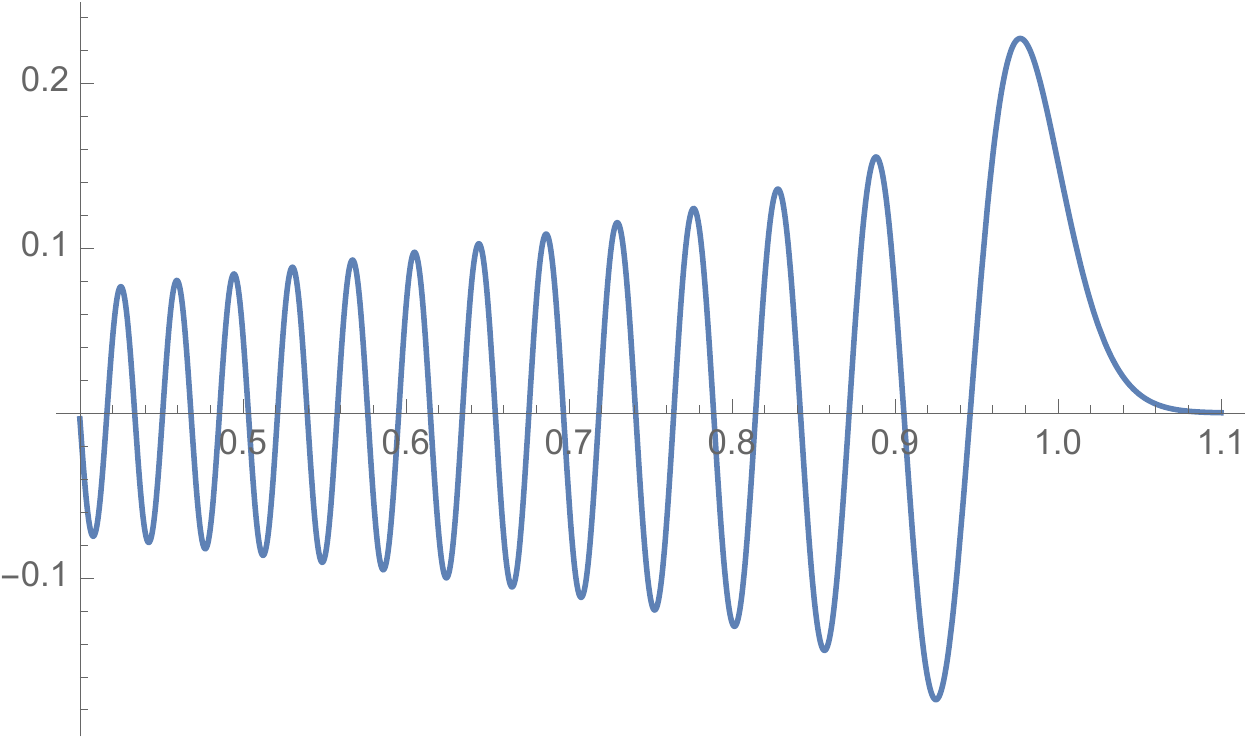}
    \caption{A plot of $(2\pi \hbar)^d H(x,\xi)^{-d/2} W_{\hbar, E}(x,\xi)$ as a function of the radial variable $\rho=\sqrt{2H(x,\xi)}$ in phase space when $E=1/2$ and $\hbar\approx 0.01$. The Airy behavior with peak  $\hbar^{1/3}\approx .22$ near the energy surface $\rho=1=\sqrt{2E}$ and the oscillatory behavior with amplitude $\hbar^{1/2}\approx .1$ in the `bulk' $H(x,\xi)<2E=1$ are clearly visible.} \label{fig:individual-wigner}
\end{figure} 

 Here, $\mathrm{Ai}$ is the Airy function (see Section \ref{AIRYAPP}).
These behaviors are illustrated in Figure \ref{fig:individual-wigner} (see also Section \ref{INDIVIDUALSECT}). In this continuation, we study `Weyl sums' \eqref{Whfdelta} of Wigner distributions over a variety of spectral windows and describe the impact of the different spectral localizations on the scaling asymptotics of the corresponding Wigner distributions in different phase space regions.  Let us begin by introducing the three types of spectral localization we are studying and the interfaces in each type. \bigskip
 
   \begin{itemize}
   
   \item (i)  $\hbar$-localized Weyl sums over eigenvalues in an $\hbar$-window  $E_N(\hbar) \in  [E - a \hbar, E + b \hbar] $ of width $O(\hbar)$. More generally we consider smoothed Weyl sums $W_{\hbar, E, f} $  with weights $f(\hbar^{-1}(E_N(\hbar) - E))$; see \eqref{UhfDEF} for such $\hbar$-energy localization. This is the scale of individual spectral projections but is substantially more general than the results of \cite{HZ19}. The scaling and asymptotics are in Theorem \ref{ELEVELLOC}. For general \Sch operators, $\hbar$- localization around a single energy level leads to expansions in terms of periodic orbits. Since all orbits of the classical isotropic oscillator are periodic, the asymptotics may be stated without reference to them. The generalization to all \Sch operators  will be studied in a future article. \bigskip


   \item (ii)  Airy-type $\hbar^{2/3}$-spectrally localized Weyl sums $W_{\hbar, f, 2/3}(x, \xi)$  
 over eigenvalues in a window $[E - a \hbar^{2/3}, E + a \hbar^{2/3}]$ of width $O(\hbar^{2/3})$. See Definition \ref{INTERDEF} for the precise definition. The levelset $\Sigma_E$ is viewed as the interface. The scaling asymptotics of its Wigner distribution across the interface are given in Theorems \ref{RSCOR} and \ref{SHARPh23INTER}. To our knowledge, this scaling has not previously been considered in spectral asymptotics.
\bigskip

   \item (iii) Bulk Weyl sums  $\sum_{N: \hbar(N +\frac{d}{2}) \in [E_1,E_2]} W_{\hbar, E_N(\hbar)}(x, \xi)$ over energies in an $\hbar$-independent  `window'  $[E_1, E_2]$ of eigenvalues;   this `bulk' Weyl sum runs over $\simeq \hbar^{-1}$ distinct eigenvalues; See Definition \ref{BULKDEF}. We are mainly interested in its scaling asymptotics around the interface $\Sigma_{E_2}$ (see Theorem \ref{BULKSCALINGCOR}). However, we also prove that the Wigner distribution  approximates the indicator function of the shell $\{E_1 \leq H \leq E_2\} \subset T^* \R^d$ (see Proposition \ref{pp:PBK-leading}). As far as we know, this is also a new result and many details are rather subtle because of oscillations inside the energy shell. Indeed, the results of \cite{HZ19} show that the indvidual terms in the sum grow like $W_{\hbar, E_N(\hbar)}(x,\xi)\simeq \hbar^{-d+1/2}$ when $H(x,\xi)\in (E_1,E_2).$ Proposition \ref{pp:PBK-leading}, in contrast, shows although the bulk Weyl sums have $\simeq \hbar^{-1}$ such terms, their sum has size $\hbar^{-d}$, implying significant cancellation. \bigskip

   \end{itemize}

  We are particularly interested in `interface asymptotics' of the Wigner-Weyl distributions $W_{\hbar, f, \delta(\hbar)}$ (see \eqref{Whfdelta}) around the edge (i.e. boundary) of the spectral interval when $(x, \xi)$ is near  the corresponding classical energy surface $\Sigma_E$.  Such edges occur when $f$ is discontinuous, e.g. the indicator function of an interval. In other words, we integrate the empirical measures \eqref{EMPDEF} below over an interval rather than against a Schwartz test function.  At the interface, there is an abrupt change in the asymptotics with a conjecturally universal shape. Theorem \ref{ELEVELLOC} gives the  shape of the interface for $\hbar$-localized sums, Theorem \ref{RSCOR} gives the shape for $\hbar^{2/3}$ localized sums and Theorem \ref{BULKSCALINGCOR} gives results on the bulk sums.   

Our results concern asymptotics of integrals of various types of test functions against  the weighted empirical measures,
\begin{equation} \label{EMPDEF} d\mu_{\hbar}^{(x, \xi)}(\tau): =
\sum_{N =0}^{\infty} W_{\hbar, E_N(\hbar)}(x, \xi) \delta_{E_N(\hbar)}(\tau),
\end{equation} and of  recentered and rescaled versions of these measures
 (see \eqref{mu2/3}  below). A  key property of  Wigner distributions of eigenspace projections \eqref{WIGNERDEF1} is  that the measures \eqref{EMPDEF} are  signed, reflecting the fact that Wigner distributions take both positive and negative values, and are  of infinite mass:

\begin{prop}\label{INFINITEPROP} The signed measures  \eqref{EMPDEF} 
are of infinite mass (total variation norm). On the other hand,  the mass of \eqref{EMPDEF} is finite on any one-sided interval of the form, $[-\infty, \tau]$.
Also,
$\int_{\R} d\mu_{\hbar}^{(x, \xi)}  = 1$ for all $(x, \xi). $\end{prop} 
 Moreover,  the $L^2$ norms of the  terms $W_{\hbar, E_N(\hbar)}$ grows in $N$ like $N^{\frac{d-1}{2}}$ (see Section \ref{NORMSECT}). Hence, the measures \eqref{EMPDEF} are highly oscillatory and the summands can be very large. This makes it difficult to study integrals of \eqref{EMPDEF} over intervals using Tauberian arguments. Instead we rely on the special spectral properties of \eqref{Hh}, encapsulated in Lemma \ref{NICEFORM} below.

   
\begin{figure}
\begin{center}
  \includegraphics[width=.5 \textwidth]{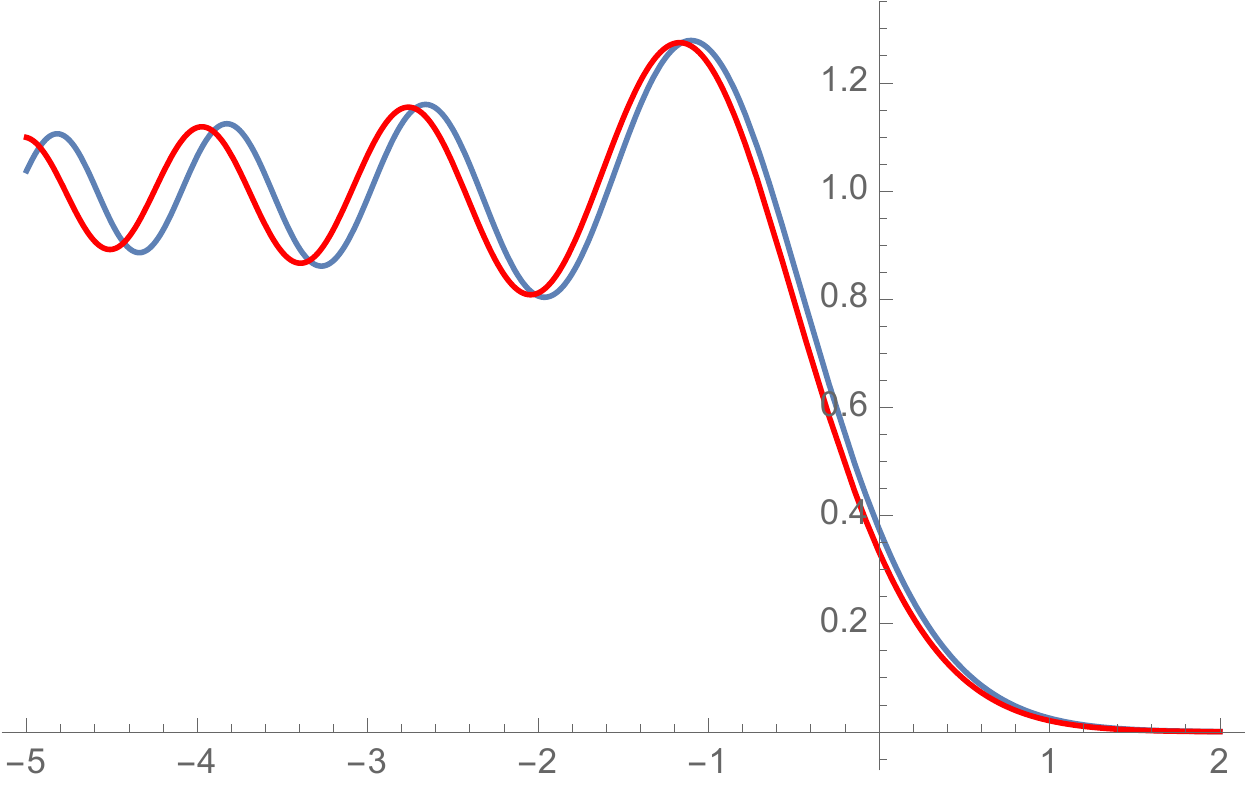} 
\end{center}
\caption{Plot with $\hbar\approx 0.02,\,E=1/2$ of scaled bulk Wigner-Weyl sum $(2\pi\hbar)^d W_{\hbar, [0,E]}(x,\xi)$ when $H(x,\xi)=E+u(\hbar/2E)^{2/3}$ as a function of $u$ (blue) against its integrated Airy limit $\int_0^\infty \mathrm {Ai}(\lambda + u/E)d\lambda$ (red) from Theorem \ref{BULKSCALINGCOR}.}
\label{fig:bulk-Wigner-Airy}
\end{figure}   

While the results of \cite{HZ19} are very special to the isotropic oscillator, many of the results of this article should be universal among  \Sch operators, at least for energy levels that are not critical values of the classical Hamiltonian.  We plan to study general \Sch operators, and the universality of scaling asymptotics, in a future article. We begin  with the case of  the isotropic oscillator to determine the form of the laws, which would be much more complicated to prove in the general case.




%

\subsection{\label{hSCSECT} Interior asymptotics for  $\hbar$-localized Weyl sums}
The first result we present pertains to the $\hbar$-spectrally localized Weyl sums of type (i), defined by taking $\delta(\hbar)=\hbar$ in \eqref{Whfdelta}:
\begin{equation}
W_{\hbar, E, f}(x, \xi) :=\sum_N f(\hbar^{-1}(E-E_N(\hbar))) W_{\hbar, E_N(\hbar)}(x,\xi),\qquad f \in \scal(\R).\label{UhfDEF}
\end{equation}

\begin{theo} \label{ELEVELLOC} Fix $E>0$, and let $W_{\hbar, E, f} $ be the Wigner distribution as in \eqref{UhfDEF} with $f$ an even Schwartz function. If $H(x,\xi)>E,$ then $W_{\hbar, E,f}(x,\xi)=O(\hbar^\infty).$ In contrast, when $0<H(x, \xi) < E$, set $H_E:=H(x,\xi)/E$ and define
\begin{align*}
t_{+,\pm,k}&:= 4\pi k \pm 2\cos^{-1}\lr{H_E^{1/2}},\quad t_{-,\pm,k}:= 4\pi \lr{k+\frac{1}{2}} \pm 2\cos^{-1}\lr{H_E^{1/2}},\qquad k \in \Z.
\end{align*}
Fix any $\delta>0.$ Then
\[
W_{\hbar, f, E}(x,\xi)=\frac{\hbar^{-d+1}\lr{1+O_\delta(\hbar^{1-\delta})}}{(2E)^{1/2}(2\pi)^dH_E^{d/2}(H_E^{-1}-1)^{1/4}}~\sum_{\pm_1,\pm_2\in \set{+,-}} \frac{e^{\pm_2 i\lr{\frac{\pi}{4}-\frac{4E}{\hbar}}}}{\lr{\pm_1}^d}\sum_{k\in \Z} \widehat{f}(t_{\pm_1,\pm_2,k})e^{\frac{iE}{\hbar}t_{\pm_1,\pm_2,k}},
\]
where the notation $O_\delta$ means the implicit constant depends on $\delta.$
\end{theo}
\noindent Note that there are potentially an infinite number of `critical points' in the support of $\hat{f}$. We prove Theorem \ref{ELEVELLOC} in Section \ref{WIGNERINTRO}.

\subsection{\label{SECT23} Interface asymptotics for smooth $\hbar^{2/3}$-localized Weyl sums}
We now consider less spectrally localized Wigner distributions \eqref{Whfdelta} but that are both spectrally localized and phase-space localized on the scale $\delta(\hbar)= \hbar^{2/3}$. They are mainly relevant when we study interface behavior around $\Sigma_E$ of Weyl sums. 

\begin{defn}\label{INTERDEF} Let $H(x, \xi) = (\norm{x}^2 + \norm{\xi}^2)/2$, and assume that $(x, \xi)$ satisfies
  \begin{equation}
H(x,\xi) =  E + u\lr{\hbar/2E}^{2/3}.\label{E:near-shell}
\end{equation}
Let  $\delta(h) = \hbar^{2/3}$ in \eqref{Whfdelta} and define the interface-localized Wigner distributions by
$$\begin{array}{lll} W_{\hbar, f, 2/3}(x, \xi): & = &   \sum_{N} f(h^{-2/3} (E-E_N(\hbar)))W_{ \hbar, E_N}(x, \xi) \;\; \end{array} $$

\end{defn}
\begin{theo} \label{RSCOR}
Assume that $(x, \xi)$ satisfies \eqref{E:near-shell} with $\abs{u}<\hbar^{-2/3}.$ Fix a Schwartz function $f \in \scal(\R)$ with compactly supported Fourier transform. Then
\[W_{\hbar, f, 2/3}(x, \xi)  = (2\pi \hbar)^{-d} I_0(u;f,E)~+~O((1+\abs{u})\hbar^{-d+2/3}),\]
where
\[I_0(u; f, E) = \int_{\R} f(-\lambda/C_E) \mathrm{Ai}\lr{\lambda + \frac{u}{E}} d\lambda,\qquad C_E=(E/4)^{1/3}.\]
More generally, there is an asymptotic expansion
\[W_{\hbar, f, 2/3}(x, \xi) ~\simeq~(2\pi\hbar)^d\sum_{m\geq 0}\hbar^{2m/3}I_m\lr{u;f,E}\]
in ascending powers of $\hbar^{2/3}$ where $I_m(u;f,E)$ are uniformly bounded when $u$ stays in a compact subset of $\R.$
\end{theo}

The calculations show that the results are valid with far less stringent conditions on $f$ than $f \in \scal(\R)$ and $\widehat{f}\in C_0^\infty$. To obtain a finite expansion and remainder it is sufficient that $\int_{\R} |\widehat{f}(t)| |t|^k dt  < \infty$ for all $k.$ It is not necessary that $\hat{f} \in C^k$ for any $k >0$. Theorem \ref{RSCOR} is proved in Section \ref{RSCOR-pf}. 

\subsection{Sharp $\hbar^{2/3}$-localized Weyl sums}
Next we consider the sums of Definition \ref{INTERDEF} when $f$ 
is the indicator function of a spectral interval, 
$$f = {\bf 1}_{[ \lambda_- ,\lambda_+]}.$$
Equivalently, we fix positive integers $n_{_\pm}$ such that
\[\lambda_{\pm}=\hbar^{1/3}n_{\pm}~~\text{are bounded},\]  
and consider the corresponding Wigner-Weyl sums $ W_{\hbar, f, 2/3}(x, \xi)$ of Definition \ref{INTERDEF}:
\begin{equation} \label{W23EDEF} W_{2/3,E,\lambda_{\pm}}(x,\xi):=\sum_{N:\, \lambda_- \hbar^{2/3}\leq E_N(\hbar)-E< \lambda_+\hbar^{2/3}}W_{\hbar, E_N(\hbar)}(x,\xi)=\sum_{N=N(E,\hbar)+n_-}^{N(E,\hbar)+n_+-1}W_{\hbar, E_N(\hbar)}(x,\xi)
,\end{equation}
where $N(E,\hbar)=E/\hbar - d/2$. Thus, the sums run over spectral intervals of size $\simeq \hbar^{2/3}$ centered at a fix $E>0$ and consist of a sum of $\simeq \hbar^{-1/3}$ Wigner functions for spectral projections of individual eigenspaces.  The following extends Theorem \ref{RSCOR} to sharp Weyl sums at the cost of only giving a 1-term expansion plus remainder.

\begin{theo} \label{SHARPh23INTER} Assume that $(x, \xi)$ satisfies $\lr{\norm{x}^2+\norm{\xi}^2}/2 =E+ u\lr{\hbar/2E}^{2/3}$ with $\abs{u}<\hbar^{-2/3}.$ Then, \[W_{2/3,E,\lambda_{\pm}}(x,\xi)=(2\pi\hbar)^{-d}C_E\int_{-\lambda_+}^{-\lambda_-}\Ai\lr{\frac{u}{E}+\lambda C_E} d\lambda + O\lr{\hbar^{-d+1/3-\delta}+(1+\abs{u})\hbar^{-d+2/3-\delta}},\]
where $ C_E = (E/4)^{1/3}.$
\end{theo}
\noindent Theorem \ref{SHARPh23INTER} is proved in Section \ref{2/3sumSect}. It can be rephrased in terms of  weighted empirical measures

\begin{equation} \label{mu2/3} d \mu^{u, E, \frac{2}{3}}_{\hbar} := \hbar^d\; \sum_{N} W_{ \hbar, E_N(\hbar)}\lr{ E + u\lr{\hbar/2E}^{2/3}} \delta_{[ \hbar^{-2/3} (E- E_N(\hbar) )]}.\end{equation} 
 obtained by centering and scaling the family \eqref{EMPDEF}. Thus, for  $(x, \xi)$ satisfying $\lr{\norm{x}^2+\norm{\xi}^2}/2 = E + u\lr{\frac{\hbar}{2E}}^{2/3},$ and for $f \in \scal(\R)$,
$$W_{\hbar, f, 2/3}(x, \xi): = \hbar^{-d} \int_{\R} f(\tau) d \mu^{u, E, \frac{2}{3}}_{\hbar}(\tau), \;\;\; W_{2/3,E,\lambda_{\pm}}(x,\xi) =   \hbar^{-d} \int_{\lambda_-}^{\lambda^+}  d \mu^{u, E, \frac{2}{3}}_{\hbar}(\tau).$$

\subsection{\label{BULKSECT} Bulk sums}
   
We next consider Weyl sums of eigenspace projections corresponding to an energy shell (or window) $[E_1, E_2]$. We consider both sharp and smoothed sums.

\begin{defn} \label{BULKDEF} Define the `bulk' Wigner distributions for an $\hbar$-independent energy window $[E_1, E_2]$ by
\begin{equation} \label{WE} W_{\hbar,[E_1, E_2]}(x, \xi): \sum_{N: E_N(\hbar) \in [E_1,E_2]} W_{\hbar, E_N(\hbar)}(x, \xi). \end{equation} 
More generally for $f \in C_b(\R)$ define
\begin{equation} \label{WEf} W_{\hbar, f}(x, \xi):= \sum_{N=1}^{\infty}
f(\hbar(N +d/2)) \; W_{\hbar, E_N(\hbar)}(x, \xi). \end{equation} 
\end{defn}
\noindent Our first result about the bulk Weyl sums concerns the smoothed Weyl sums $W_{\hbar, f}.$ 

\begin{prop} \label{smoothedbulk}For $f \in \scal(\R)$ with $\hat{f} \in C_0^{\infty}$,  $W_{\hbar, f}(x, \xi)$ admits a complete asymptotic expansion as $\hbar \to 0$ 
of the form,
$$\left\{ \begin{array}{lll} W_{\hbar, f}(x,\xi) & \simeq &  (\pi \hbar)^{-d} \sum_{j = 0}^{\infty} c_{j, f, H}(x, \xi) \hbar^j,\; \rm{with} \\&&\\c_{0, f, H} (x, \xi) 
& = & f(H(x, \xi)) = \int_{\R} \hat{f}(t) e^{i t H(x,\xi)} dt.

\end{array}\right. $$
In general $c_{k, f, H}(x, \xi)$ is a distribution of finite order on $f$ supported
at the point $(x, \xi)$.

\end{prop}
\noindent The proof is given in Section \ref{BULKfSECT} and merely involves
Taylor expansion of the phase. 

\subsection{Interior/exterior asymptotics for bulk Weyl sums of Definition \ref{BULKDEF}}
From Proposition \ref{smoothedbulk}, it is evident that the behavior of $W_{\hbar, [E_1,E_2]}(x,\xi)$ depends on whether $H(x, \xi) \in (E_1, E_2)$ or $H(x, \xi) \notin [E_1, E_2]$. Some of this dependence is captured in the following result. 
\begin{prop}
\label{pp:PBK-leading}  We have,
\[ 
W_{\hbar, [E_1, E_2]}(x, \xi) = \left\{
\begin{array}{ll} (i)\; 
(2\pi\hbar)^{-d} (1+O(\hbar^{1/2})), & 
 H(x, \xi) \in (E_1, E_2),\\ &\\ 
(ii) \;  O(\hbar^{-d +1/2}), & H(x, \xi) < E_1, \\&\\ 
 (iii) \;  O(\hbar^{\infty}), & H(x, \xi) > E_2
\end{array} \right.
\]
\end{prop}

The two `sides' $0 < H(x, \xi) < E_1$ and $H(x, \xi) > E_2$ behave differently because the Wigner distributions have slowly decaying tails inside an energy ball but are exponentially decaying outside of it (see Section \ref{INDIVIDUALSECT}). If we write $W_{\hbar, [E_1, E_2]}(x, \xi)  = W_{\hbar, [0, E_2]}(x, \xi) -W_{\hbar, [0, E_1]}(x, \xi) $, we see that the two  cases with $H(x, \xi) > E_1$ are covered by results for $W_{\hbar, [0, E]}$ with $E = E_1 $ or $E = E_2$. When $H(x, \xi) < E_1$, then both terms of $W_{\hbar, [0, E_2]}(x, \xi) -W_{\hbar, [0, E_1]}(x, \xi) $ have the order of magnitude $\hbar^{-d}$ and the asymptotics reflect the cancellation between the terms. The boundary case where $H(x, \xi) = E_1,$ or $H(x, \xi)= E_2$ is special and is given in Theorem \ref{RSCOR}.

\subsection{Interface asymptotics for bulk  Weyl sums of Definition 
\ref{BULKDEF}}

Our final result concerns the asymptotics of $W_{\hbar, [E_1, E_2]}(x, \xi)$ in $\hbar^{2/3}$-tubes around the `interface' $H(x, \xi) = E_2$.  Again, it is sufficient to consider intervals $[0, E]$. It is at least intuitively clear that the interface asymptotics will depend only on the individual eigenspace projections with eigenvalues in an $\hbar^{2/3}$-interval around the energy level $E$, and since they add to $1$ away from the boundary point, one may expect the  asymptotics to be similar to the interface asymptotics  for individual eigenspace projections in \cite{HZ19}. 

\begin{theo}\label{BULKSCALINGCOR} 
Assume that $(x, \xi)$ satisfies $\lr{\norm{x}^2+\norm{\xi}^2}/2 = E + u\lr{\hbar/2E}^{2/3}$ with $\abs{u}<\hbar^{-2/3}.$ Then, for any $\epsilon>0$ 
\[
W_{\hbar, [0, E]}(x, \xi)  =   \lr{2\pi \hbar}^{-d}  \left[\int_0^{\infty} \Ai\lr{\frac{u}{E}+\lambda} d\lambda +O(\hbar^{1/3-\epsilon}\abs{u}^{1/2})+ O(\abs{u}^{5/2}\hbar^{2/3-\epsilon})\right],
\]
where the implicit constant depends only on $d,\epsilon.$
\end{theo}
The Airy scaling the Wigner function is illustrated in Figure \ref{fig:bulk-Wigner-Airy}.


\subsection{\label{HEURISTICSECT} Heuristics} In Sections \ref{NORMSECT}  - \ref{INDIVIDUALSECT}, we review the results on $L^2$ norms and pointwise asymptotics of Wigner distribution of eigenspace  projections \eqref{PiDEF}. Wigner distributions are normalized so that the Wigner distribution of an $L^2$ normalized eigenfunction has $L^2$ norm $1$ in $T^*\R^d$. Due to the  multiplicity $N^{d-1}$ of eigenspaces \eqref{VhE}, the $L^2$ norm of $W_{\hbar, E_N(\hbar)}$ is  of order $N^{\frac{d-1}{2}}$.

  In the main results, we sum over windows of eigenvalues, e.g. $\lambda_- \hbar^{2/3}\leq E-E_N(\hbar)< \lambda_+ \hbar^{2/3}$ \eqref{W23EDEF}, resp. $E_N(\hbar) \in [0,E]$ in \eqref{BULKDEF}. Inevitably, the asymptotics are joint in $(\hbar, N)$. As $\hbar \downarrow 0$, the number of $N$ contributing to the sum  grows at the rate $\hbar^{-\frac{1}{3}}$, resp. $\hbar^{-1}$. Due to the $N$-dependence of the $L^2$ norm, terms with higher $N$ have norms of higher weight in $N$ than those of small $N$ but the precise size of the contribution depends on the position of $(x, \xi)$ relative to the interface $\{H = E\}$ and of course the relation \eqref{ENh}.
  
As reviewed in Section \ref{INDIVIDUALSECT},   $W_{\hbar, E_N(\hbar)}(x, \xi)$ peaks when $H(x, \xi) = E_N(\hbar)$, exponentially decays in $\hbar$ when $H(x, \xi) > E_N(\hbar)$ and has slowly decaying tails inside the energy ball $\{H < E_N(\hbar)\}$, which fall into three regimes: (i) Bessel near $0$, (ii) oscillatory or trigonometric in the bulk, and (iii) Airy near $\{H = E\}.$ In terms of $N$, when \eqref{ENh} holds, and $H(x, \xi)  < E_N(\hbar)$, then $W_{\hbar, E_N(\hbar)}(x, \xi) \simeq \hbar^{-d + 1/2} \simeq N^{d -1/2}$. Near the peak point, when $H(x,\xi)-E_N(\hbar)\approx \hbar^{2/3}$, we have in contrast $W_{\hbar, E_N(\hbar)}(x, \xi) \simeq \hbar^{-d + 1/3} \simeq_E N^{d - 1/3}.$
 
 It follows that the terms with a high value of $N$ and with $ E_N(\hbar) \geq H(x, \xi)$  in  \eqref{EMPDEF}  contribute high weights. There are an infinite number of such terms, and so \eqref{EMPDEF} is a signed measure of infinite mass (as stated in Proposition \ref{INFINITEPROP}.) This is why we mainly consider the restriction of the measures \eqref{EMPDEF} to compact intervals. See Section \ref{EMPSECT} for more on these measures.

\subsection{Discussion of Results}
Proposition  \ref{pp:PBK-leading} shows that when properly normalized (by multiplying by $\hbar^{d}$), the Wigner distribution is asymptotically equal to $1$ in the allowed region $\{E_1 \leq H \leq E_2\}$,  and asymptotically equal to $0$ in the complementary  forbidden region,  where $H(x, \xi) = \half (||\xi||^2 + ||x||^2)$ is the classical Hamiltonian; this proves a conjecture in \cite{JZ}. The main question is the shape of the interface interpolating between the values $0$ and $1$. The result (Theorem \ref{RSCOR}) is that the scaled interface shape around $\{H = E_2\}$  is the graph of the exponentially decaying side  of the Airy function  from the region around its global maximum (slightly to the left of $0$) to the right to $+\infty$. The addition of many terms in the Weyl sum fills in the region $\{E_1 < H < E_2\}$ to an asymptotically constant function  (see Figure \ref{fig:bulk-Wigner-Airy}). 
As the figure shows, there  is an asymmetry between the upper endpoint $E_2$ and the lower endpoint $E_1$, reflecting the asymmetric behavior of the Airy function on positive/negative axis. On the positive axis, $\mathrm{Ai}(x) \sim C x^{-1/4} e^{- 2x^{3/2}/3}, \;\; x \in \R_+, x\to \infty$, i.e. the decay is almost as fast as in the complex domain \cite{ZZ17}. However, $\mathrm{Ai}(-x) \sim C x^{-1/4} \sin\left(2 x^{3/2}/3 + \pi/4\right)$ when $x \to \infty$. As a result, there is a slowly decaying oscillating tail of $W_{\hbar, E_N(\hbar))}(x, \xi)$ with $E_N(\hbar) \simeq E$ for $(x, \xi)$ inside the energy surface  $\Sigma_E$ but a fast exponential decay outside of it. The oscillatory behavior inside the energy surface is in marked contrast to the Erf interface behavior in the complex domain studied in \cite{ZZ17} (see Section \ref{PRIOR}).
   
It is interesting to compare the results of Theorems \ref{RSCOR}-\ref{SHARPh23INTER}  and Theorem \ref{BULKSCALINGCOR}. In both theorems, we assume  that $(x, \xi)$ satisfies $\lr{\norm{x}^2+\norm{\xi}^2}/2=  E + u\lr{\hbar/2E}^{2/3}.$ For sharp $\hbar^{2/3}$ localization to an interval $[E+\lambda_-\hbar^{2/3},E+\lambda_+\hbar^{2/3}],$ we get   \[W_{2/3,E,\lambda_{\pm}}(x,\xi)=(2\pi\hbar)^{-d}C_E\int_{-\lambda_+}^{-\lambda_-}\Ai\lr{\frac{u}{E}+\lambda C_E} d\lambda + O(\hbar^{-d+1/3-\delta}),\qquad C_E = (E/4)^{1/3},\]
while at the interface of the bulk sums we get essentially the same asymptotics,
\[
W_{\hbar, [0, E]}(x, \xi)  =   \lr{2\pi \hbar}^{-d}  \int_0^{\infty} \Ai\lr{\frac{u}{E}+\tau} d\tau + O((1+\abs{u})\hbar^{-d+1/3-\epsilon}).
\]
One may understand this coincidence by decomposing  the bulk  Weyl sum,     \begin{equation} \label{WEDECintro} W_{\hbar, [0,E]}(x, \xi)=   \sum_{N: \hbar(N +\frac{d}{2}) \in [0, E - \delta(\hbar)]} W_{\hbar, E_N(\hbar)}(x, \xi) +\sum_{N: \hbar(N +\frac{d}{2}) \in [E - \delta(\hbar), E]} W_{\hbar, E_N(\hbar)}(x, \xi), \end{equation} 
with $\delta(\hbar) \simeq \hbar^{2/3}$, into terms with energy levels either sufficiently close or sufficiently well-separated from the boundary level $E$. One then shows that the terms far from the interface do not make a principal order contribution.  Intuitively, the reason is that  the summands $W_{\hbar, E_N(\hbar)}$ with $E_N(\hbar) \simeq a < E$ are exponentially decaying to the right of $H(x,\xi) = a$, or better to the right of the maximum point at $\hbar^{2/3} (a - \zeta_0)$ where $\zeta_0$ is the first critical point of $\rm{Ai}(x)$, hence contribute  little to the asymptotics for the given $(x, \xi)$. The further that $E_N(\hbar) \simeq a$ moves left of $E$,  the more exponentially decaying  $W_{\hbar, E_N(\hbar)}(x, \xi)$ becomes. Hence it is only at the edge of the interval $[0, E - C \hbar^{2/3}]$ that the terms of the first sum can contribute non-trivially to the asymptotics. Each of the terms with $E_N(\hbar) \simeq E + C \hbar^{2/3} $ is in fact positive since the largest zero of $\rm{Ai}(x)$ is approximately  $ - 2.3381.$ Each is of size $h^{-d+1/3}$ (see Section \ref{INDIVIDUALSECT})  and there are $\hbar^{-1/3}$ such terms, so the sum over the interval $[E - C \hbar^{2/3}, E]$ does account for the order of magnitude. This heuristic discussion ignores the fact that the terms are signed and that cancellation is needed for the terms with $E_N(\hbar) \leq E - C \hbar^{2/3}$ to obtain the correct order of magnitude.

\subsection{\label{Outline}Outline of proofs } 

As in \cite{HZ19}, we express Wigner functions $W_{\hbar, E_N(\hbar)}$ of projections $\Pi_{\hbar, E_N(\hbar)}$ onto the $N^{th}$ eigenspace of $\widehat{H}_\hbar$ as Fourier coefficients of the Wigner function $\ucal_{\hbar}(t)$ of the Propagator $U_{\hbar}(t)= e^{- i \frac{t}{\hbar} \hat{H}_{\hbar}}$. To make this precise, notice that on the level of Schwartz kernels,
\begin{equation}\label{PROJUT}
\Pi_{\hbar, E_N(\hbar)}(x,y)=\int_{-\pi}^{\pi} U_\hbar(t-i\epsilon,x,y) e^{-\ihbar
(t-i\epsilon) E_N(\hbar)} \frac{dt}{2\pi}
\end{equation}
is a Fourier coefficient of the boundary value of the function $U_\hbar(t,x,\xi)$ of $t$, which is holomorphic for $t$ in the lower half-plane. The Wigner function of the propagator $\ucal_\hbar(t,x,\xi)$ is defined as
\begin{equation} \label{WIGU} \ucal_\hbar(t, x,\xi) := \int e^{- \ihbar 2 \pi  \xi p} U_{\hbar}\lr{t,x + \frac{p}{2}, x - \frac{p}{2}}dp. \end{equation}
Since the map from a kernel to its Wigner trasnform si linear, we can take the Wigner transform of the identities above between $\Pi_{\hbar, E_N(\hbar)}$ and $U_\hbar(t,x,y)$ to see that $W_{\hbar, E_N(\hbar)}$ is essentially a Fourier coefficient of $\ucal_\hbar(t):$
\begin{equation}
W_{\hbar, E_N(\hbar)}(x,\xi)=\int_{-\pi}^\pi e^{itE_N(\hbar)/\hbar} \ucal_\hbar(t,x,\xi)\frac{dt}{2\pi}.\label{E:Fourier-Wigner}
\end{equation}
This expression is profitably combined with the following exact formula for the Wigner distribution $\ucal_{\hbar}(t, x, \xi) $ of the propagator.
\begin{proposition}[Proposition 1.9 in \cite{HZ19}]\label{P:Wigner-prop} We have,
$$\ucal_{\hbar}(t, x, \xi) =(2\pi\hbar\cos((t+i0)/2))^{-d} e^{-2iH\tan(t/2)/\hbar}, \quad H=H(x,\xi)=\frac{1}{2}\lr{\norm{x}^2+\norm{\xi}^2}.$$
\end{proposition}
\noindent The Wigner distribution $\ucal_\hbar(t,x,\xi)$ is well-defined as a distribution (see \cite{HZ19}). For any $f\in \mathcal S(\R)$ and any $\delta(\hbar)>0$ we may use Fourier inversion to write
\begin{align}
\notag W_{\hbar, f, E}(x,\xi)&~=~\sum_{N\geq 0} f\lr{\delta(\hbar)^{-1}\lr{E-E_N(\hbar)}}W_{\hbar, E_N(\hbar)}(x,\xi)\\
&~=~\int_{\R}\widehat{f}(t) e^{itE\delta(\hbar)^{-1}}\ucal_\hbar(t\hbar\delta(\hbar)^{-1}, x,\xi)\frac{dt}{2\pi}.\label{E:smoothed-sum}
\end{align}
To obtain the results on smoothed Wigner-Weyl sums (Theorems \ref{ELEVELLOC} and \ref{RSCOR} as well as Proposition \ref{smoothedbulk}), where $f$ is Schwartz, we use this formula with $\delta(\hbar)=\hbar,\hbar,1$ and simply Taylor expand the phase and amplitude from Proposition \ref{P:Wigner-prop}. A key point is that $\ucal_\hbar(t,x,\xi)$ is a smooth function away from $t\in \pi \Z$ and that the contribution to the integral in \eqref{E:smoothed-sum} from a neighborhood of any of these points is $O(\hbar^\infty).$ Since $\ucal_\hbar(t)$ is periodic, it suffices to check this near $t=\pi,$ which is the content of a result from \cite{HZ19}. To state it, fix $ \delta>0$ and define a sequence of smooth of cut-off functions
\[\chi_{\hbar}:\R\gives [0,1],\quad \chi_{\hbar}(t)=
\begin{cases}
  1,&\quad \abs{t-\pi}<\hbar^\delta\\
 0, &\quad \abs{t-\pi} > 2\hbar^\delta
\end{cases}
.\]
\begin{lemma}[Lemma 3.1 in \cite{HZ19}]\label{LOCLEM}
Fix $\epsilon>0$. Uniformly over $(x,\xi)$ with $H(x,\xi)>\epsilon,$ the localized version
\begin{equation}
  \label{E:smoothed-sum-2}
  \int_{\R} e^{-itE/\hbar} \chi_\hbar(t)\widehat{f}(t) \ucal_\hbar(t,x,\xi)\frac{dt}{2\pi}
\end{equation}
of \eqref{E:smoothed-sum} is $O(\hbar^\infty).$
\end{lemma}
The idea behind this lemma is that $\ucal_\hbar(t)$ is not locally $L^1$ near $t\in \pi\Z,$ but is nonetheless well-defined as a tempered distribution in the sense that the integral
        \begin{equation}\label{ucalhf}  
\ucal_{\hbar, f}(x, \xi) : =\int_{\R} \hat{f}(t)  \ucal_{\hbar}(t, x, \xi)dt.  \end{equation}
is well-defined for $f \in \scal(\R).$  In fact, $\ucal_{\hbar}(t+\pi, x, \xi)$ resembles the position space propagator \eqref{E:Mehler} with $y = \xi$ except that its phase lacks the term $- \frac{2}{\cos t/2} \langle x, \xi \rangle $. Note that for small $t$ the phase of \eqref{E:Mehler} is essentially $\frac{x^2 + y^2}{2 t} - \frac{1}{t}\langle x, y \rangle = |x - y|^2/2t$ and that as $t \to 0$ this kernel weakly approaches $\delta(x - y)$. Similarly,  $\ucal_{\hbar} (t, x, \xi) \to \delta_0(x, \xi)$ as $t \to  \pi$, the Dirac distribution at the point $(0,0) \in T^* \R^n$. Indeed, at $t=\pi$ it is the Wigner distribution of $\delta(x + y)$, which is $\int_{\R^n} \delta(x - \frac{v}{2} + x + \frac{v}{2}) e^{- i \langle v, \xi \rangle} dv = \delta(2x) \delta(\xi). $ Since $\ucal_\hbar$ is a locally $L^1$ function at times $t \notin \pi \Z$ and is a measure when $t \in \pi \Z$ it is a measure for all $t$.

To prove our sharp results when $f$ is the indicator function of an interval (Theorems \ref{SHARPh23INTER} and \ref{BULKSCALINGCOR} as well as Proposition \ref{pp:PBK-leading}), we make key use of the special properties of the isotropic oscillator. Namely, we use the special nature of its spectrum, which after rescaling by $\hbar$ and subtracting $d/2$ forms an arithmetic progression. A simple identity is useful when dealing with sums over spectral intervals rather than smooth sums.  Since we use it several times, we record it in advance:
 
  \begin{lem} \label{NICEFORM}  We have,
\begin{equation} \label{EXACT}  \sum_{N: E_1\leq E_N(\hbar) < E_2} W_{\hbar, E_N(\hbar)}(x, \xi) = \int_{S^1} \frac{e^{i N_2 t} -e^{iN_1t}}{e^{it} -1}e^{itd/2}\ucal_{\hbar}(t, x, \xi)  \frac{dt}{2\pi},
\end{equation}
where $N_i=E_i/\hbar-d/2,\, i=1,2.$ 
\end{lem}
\noindent The identity follows directly from \eqref{E:Fourier-Wigner} by noting that $E_N(\hbar)/\hbar = N+d/2$ and summing a partial geometric series. The expression $\lr{e^{i N_2 t} -e^{iN_1t}}/\lr{e^{it} -1}$ when $N_1=0$ is the Dirichlet kernel, and its analysis depends on the scale of $N_2-N_1$ (see the discussion in Section \ref{PROOFBULK1} for an explanation of which intervals in $t$ are important for analyzing \eqref{EXACT}). In several cases, we will find that the dominant contribution comes from $t\approx 0$.

\subsection{\label{PRIOR}Prior results and further questions}
As mentioned above, this article is a continuation of \cite{HZ19}, which
contains further references to the vast literature on Wigner distributions
and the isotropic Harmonic Oscillator. Most of that literature pertains to dimension 1. As also mentioned, the problems studied here can be posed for  general \Sch operators, and we expect some of the interface scaling results to be universal. 

The weights of  the  empirical measures \eqref{EMPDEF} are given by
$W_{\hbar, E_N(\hbar)}(x, \xi)$, whose size depends on the pair $(\hbar, N)$. 
A potentially interesting  option  
is to weight the terms in \eqref{EMPDEF} with functions $w_N$ of $N$, 
e.g. $N^{-\alpha}$ or the weights $(E_N(\hbar) - E)_+^{\alpha}$ of a Riesz mean, to counter-act the  growth of the Wigner weights as $N \to \infty$ and 
change the infinite measures \eqref{EMPDEF} to finite ones.

The  interface asymptotics of Theorem \ref{RSCOR} and Theorem \ref{BULKSCALINGCOR}  may seem reminiscent of the Gibbs phenomenon for Fourier series of indicator functions, where one obtains a universal $\frac{\sin t}{t}$-integral interface in configuration space (the circle) at a jump.
Generalizations to manifolds are given in \cite{PT97}.  They are of a different nature than the interface results here, although there is some common ground. Here,  the interface is due to a sharp cutoff ${\bf 1}_{[E_1, E_2]}$ in the spectral window, while the Gibbs phenomenon pertains to convergence of eigenfunction expansions of discontinuous functions of the physical space variable $x$.  In particular, the Gibbs phenomenon is essentially a physical space phenomenon while the interface phenomena studied here are essentially phase space phenomena. 

To clarify the analogy with the Gibbs phenomenon, consider the \Sch operator $\hbar D$ on the circle $S^1$, where $D =\frac{d}{i dx}$. Its eigenvalues (resp. eigenfunctions) are $E_n(\hbar) = \hbar n$ (resp.  $e^{i n x}$). For each $x$ we  define a (complex) empirical measure by $\mu_h^x : = \sum_{n =-\infty}^{\infty} e^{i n x} \delta_{\hbar n} $. This is not really analogous to \eqref{EMPDEF} because the weights are physical space functions rather than phase space functions. Then
$$\mu_h^x[-E,  E] = D_{\hbar,E}(x):=\frac{\sin  (N(\hbar, E) + \half)x}{\sin \frac{x}{2}},$$ 
where $N(\hbar, E) =\max\{n: E_n(\hbar) \leq E\} \iff n \leq \hbar^{-1} E\}$, and where $D_{\hbar, E}$ is the Dirichlet kernel. This kernel arises in the proof Lemma \ref{NICEFORM} and in \eqref{EXACT}, suggesting a connection to the Gibbs phenomenon.  In  the proof of the Gibbs phenomenon for step functions with a jump at $x = 0$, one lets $x_n = y (n + \half)^{-1}$ for some $y >0$. Then, $D_{\hbar, E}(x_{N(\hbar, E)} ) = \frac{\sin y}{y}  + O(\hbar)$. The indefinite integral $Si(y) = \int_0^{y} \frac{\sin t}{t} dt$ gives the interface asymptotics of the Gibbs phenomenon when the jump occurs at $x = 0$.

      In contrast, the problem we study for the Harmonic oscillator is the behavior of Wigner-Weyl sums over $\{n: \hbar n \leq E\}$  at the boundary $|\xi| = E$ of an energy ball $\{|\xi| \leq E\} \subset T^*S^1$. Since the energy curve projects over all of $S^1$ without singularities, there is no analogue of allowed or forbidden regions; and since $\mu_{\hbar}^x$ does not depend on 
      the momentum $\xi$, there is no interface behavior to consider at $\{|\xi|= E\}$. There could exist an analogous interface problem if one defined phase
      space measures on $T^*S^1$ analogous to Wigner distributions. However, the most natural analogues are $\delta$-functions on energy levels, about which little is left to be known.  In this 
      direction, we note that Guillemin-Sternberg \cite{GS81} studied  Weyl asymptotics in many realizations of the metaplectic representation, e.g. the \Sch representation studied here, the Bargmann-Fock representation studied in \cite{ZZ17} and in 
      a largely unexplored context of the Lagrangian Grassmannian, a generalization of $S^1$. More precisely, they  studied intertwining Fourier
      integral operators between Harmonic Oscillators (and other quadratic Hamiltonians) from $L^2(\R^d)$ to $L^2(\Lambda_d)$ where $\Lambda_d$ is the Lagrangian Grassmannian. When $d =1$, $\Lambda_d = S^1/\Z_2$. 
      It might be interesting to study the induced map from Wigner distributions to phase space distributions in the setting of $\Lambda_d$.

In view of the analogies between Harmonic oscillators and Fourier series, there should exist a direct analogue of the Gibbs phenomenon for `Fourier-Wigner' series of radial functions on $T^* \R^d$. Namely,   the Wigner distributions \eqref{WIGNERDEF1}  form an orthonormal basis of radial functions on 
   $T^*\R^d$ (see \cite{HZ19} for background).  Hence the  `Fourier-Wigner series'  expansion of the indicator function ${\bf 1}_{\{H \leq E\}}$
of an energy ball in terms of \eqref{WIGNERDEF1} must diverge across
the energy surface $\Sigma_E$ in a universal way. Almost certainly, it also diverges   at the origin, giving a Wigner analogue of the `Pinsky phenomenon.'

\section{Background}

In this section we give a rapid review of the results on Wigner distributions of individual eigenspace projections in \cite{HZ19}, so that the reader may compare them with the results on  Weyl sums of projections in this article. We also review the relation to Laguerre polynomials and  prove Proposition \ref{INFINITEPROP}.

 \subsection{\label{NORMSECT}Norms and normalizations}  Recalling the  notation in Section \ref{INTRO}, the isotropic oscillator \eqref{Hh} admits the spectral decomposition 
\begin{equation}  
L^2(\R^d, dx) = \bigoplus_{N\in \mathbb N } V_{\hbar,E_N(\hbar)}, \qquad \widehat{H}_\hbar|_{V_{h,E_N(\hbar)}}  =  \hbar (N + d/2).
\end{equation}  

The Wigner transform, which is typically viewed as a mapping
$$\wcal_{\hbar}: L^2(\R^d \times \R^d) \to L^2(T^*\R^d),$$  
in fact acts on Schwartz kernel distributions $K_{\hbar} \in \dcal'(\R^d \times \R^d)$ as follows:
\begin{equation}
 \wcal_{\hbar}(K_{\hbar})(x, \xi): = (2 \pi\hbar)^{-d} \int_{\R^d} K_{\hbar} \left( x+\frac{v}{2}, x-\frac{v}{2} \right) e^{-\frac{i}{\hbar} v \xi} \frac{dv}{(2\pi h)^d}. \label{E:WignerK}
\end{equation} 

Taking $K_\hbar$ to be an eigenspace projection $\Pi_{\hbar, E_N(\hbar)}(x,y)$ \eqref{PiDEF} gives the Wigner distributions $ W_{\hbar, E_N(\hbar)}(x, \xi)$ of individual eigenspace projections \eqref{WIGNERDEF1}.
In \cite{HZ19} it is proved that,
\begin{equation} \label{INTEGRALS}\left\{  \begin{array}{ll}(i) &  \int_{T^* \R^d} W_{\hbar, E_N(\hbar)}(x, \xi) dx d \xi = \rm{Tr} \Pi_{\hbar, E_N(\hbar)} = \dim V_{\hbar, E_N(\hbar)} = \binom{N+d-1}{d-1} \\ &  \\(ii) & 
\int_{T^* \R^d}\left| W_{\hbar, E_N(\hbar)}(x, \xi) \right|^2 dx d \xi = \rm{Tr} \Pi^2_{\hbar, E_N(\hbar)} = \dim V_{\hbar, E_N(\hbar)}=\binom{N+d-1}{d-1}\\ & \\ (iii)& \int_{T^* \R^d} W_{\hbar, E_N(\hbar)}(x, \xi) \overline{W_{\hbar, E_M(\hbar)}(x, \xi)}dx d \xi = \mathrm{Tr} \Pi_{\hbar, E_N(\hbar)}  \Pi_{\hbar, E_M(\hbar)}= 0, \; \rm{for}\; M \not=N. \end{array} \right.,\end{equation} 
In these equations, $N=\frac{E}{\hbar}-\frac{d}{2},$ and $\binom{N+d-1}{d-1}$ is the composition function of $(N,d)$ (i.e. the number of ways to write $N$ as an ordered us of $d$ non-negative integers). Thus, the sequence,
$$\{ \lr{\dim V_{\hbar, E_N(\hbar)}}^{-1/2} W_{\hbar, E_N(\hbar)}\}_{N=1}^{\infty}\subset L^2(\R^{2n}) $$
is orthonormal.

\subsection{\label{INDIVIDUALSECT}Asymptotics of individual eigenspace projection Wigner distributions}

Next we review  the results of \cite{HZ19} on pointwise asymptotics of Wigner distributions of individual eigenspace projections, which are necessary to understand the behavior of sums. In each result we fix one term by requiring that $E_N(\hbar) = E$, and then give pointwise results depending on the position of $(x, \xi)$ with respect to \eqref{SIGMAEDEF}. The first result gives Airy scaling asymptotics around the energy surface:

\begin{theo}\label{SCALINGCOR-old} 
Fix $E>0,d\geq 1$. Assume $E_N(\hbar)  = E$.
Suppose $(x,\xi)\in T^*\R^d$ satisfies
\begin{equation} \label{udef}
H(x,\xi)=E + u \lr{\frac{\hbar}{2E}}^{2/3},\qquad u \in \R,\, H(x,\xi)=\frac{\norm{x}^2+\norm{\xi}^2}{2}\end{equation}
with $\abs{u}<\hbar^{-1/3}$. \footnote {The errors blow up when $u=\hbar^{-1/3}$.} Then,
\begin{equation}
  \label{E:W-scaling}
 W_{\hbar, E_N(\hbar)}(x, \xi)=
  \begin{cases}
 \frac{2}{(2\pi \hbar)^d} \lr{\frac{\hbar}{2E}}^{1/3} \lr{\Ai(u/E) + O\lr{(1+\abs{u})^{1/4}u^2\hbar^{2/3}}},&\qquad u<0\\
 \frac{2}{(2\pi \hbar)^d} \lr{\frac{\hbar}{2E}}^{1/3} \Ai(u/E)\lr{1 + O\lr{(1+\abs{u})^{3/2}u\hbar^{2/3}}},&\qquad u>0
  \end{cases}
\end{equation}
\end{theo}

\noindent The second gives pointwise asymptotics when $(x, \xi)$ lies in the interior of the energy ball (the `bulk'):

 \begin{theo} \label{BESSELPROP} Fix $E>0$ and suppose $E_N(\hbar)=E.$ For each $(x,\xi)\in T^*\R^d$ write
\[H_E := \frac{H(x,\xi)}{E}=\frac{\norm{x}^2+\norm{\xi}^2}{2E},\qquad \nu_E:=\frac{4E}{\hbar}.\]
Fix $0<a < 1/2.$ Uniformly over $a\leq H_E \leq 1-a$, there is an asymptotic expansion,
\[W_{\hbar, E_N(\hbar)}(x, \xi)=\frac{2}{(2\pi\hbar)^d}\left[\frac{J_{d-1}(\nu_E A(H_E))}{A(H_E)^{d-1}}\alpha_0(H_E)+O\lr{ \nu_E^{-1}\abs{\frac{J_{d}(\nu_E A(H_E))}{A(H_E)^{d}}}}\right].\]
In particular, uniformly over $H_E$ in a compact subset of $(0,1),$ we find
\begin{equation}
W_{\hbar, E_N(\hbar)}(x, \xi) = (2\pi\hbar)^{-d+1/2} P_{H,E}\cos\lr{\xi_{\hbar, E,H}}+O\lr{\hbar^{-d+3/2}},\label{E:interior-cosine}
\end{equation}
where we've set
\[ \xi_{\hbar, E,H} =-\frac{\pi}{4} -\frac{2H}{\hbar}\lr{H_E^{-1}-1}^{1/2}+\frac{2E}{\hbar}\cos^{-1}\lr{H_E^{1/2}}\]
and 
\[P_{E,H}:=\lr{\pi E^{1/2}\lr{H_E^{-1}-1}^{1/4}\lr{H_E}^{d/2}}^{-1}.\]
\end{theo}

\noindent The third gives asymptotics in the `forbidden region' where $(x, \xi)$ lies in the exterior of the energy surface.

\begin{prop}\label{EXTDECAY} Suppose that $H_E=H(x, \xi)/E>1$ and let $E_N(\hbar)  =E$. Then, there exists $C_1>0$ so that
\[|W_{\hbar, E_N(\hbar)}(x, \xi)| \leq C_1 \hbar^{-d + \frac{1}{2}} e^{- \frac{2 E}{\hbar}[\sqrt{H_E^2 -H_E} - \cosh^{-1}\sqrt{H_E} ]}. \]
Moreover, as $H(x,\xi) \to \infty$, there exists $C_2>0$ so that
$$|W_{\hbar, E_N(\hbar)}(x, \xi)| \leq C_2 \hbar^{-d + \frac{1}{2}} e^{- \frac{2H(x,\xi)}{\hbar}}. $$
\end{prop}

\subsection{\label{WLSECT} Wigner-Laguerre } 

In \cite{HZ19} (see also \cite{T, T12}), it is shown that

\begin{equation}    \label{E:Wigner-sp}
    W_{\hbar, E_N(\hbar)}(x, \xi) =  \frac{(-1)^N}{(\pi \hbar)^d}
    e^{-  2H/\hbar}  L^{(d-1)}_N(4H/\hbar),\qquad H=H(x,\xi)=\frac{\abs{x}^2+\abs{\xi}^2}{2},
\end{equation}
where $L_N^{(d-1)}$ is the associated Laguerre polynomial of degree $N$ and type $d-1$. Classical asymptotics of Laguerre functions (see \cite{AT15, AT15b} for recent results and references) show that for $x$ small relative to $N$, $e^{-x/2} L^{(d-1)}_N(x)$   can be asymptotically approximated by Bessel functions. To the right, in the bulk,  the oscillatory behavior is modeled by trigonometric functions, and in a neighborhood of the extreme right by Airy functions. In this article, we are concerned with sums over $N$ of such Laguerre functions, and use semi-classical methods different from the classical ones, but this description is obviously visible in the main results of this article and in \cite{HZ19}.
 
\subsection{\label{EMPSECT} Empirical measures:    Proof of Proposition \ref{INFINITEPROP}}
First let us consider the (conditionally convergent) integral of \eqref{EMPDEF}. Letting $K_{\hbar} = \Pi_{\hbar, E_N(\hbar)}$ and noting that $\sum_{N = 0}^{\infty}\Pi_{\hbar, E_N(\hbar)}(x,y) = \delta(x - y)$, we have
\begin{align*}
\hbar^d\sum_N W_{ \hbar, E_N}(x, \xi) &=\hbar^d \wcal_{\hbar} (\sum_{N =1}^{\infty}  \Pi_{\hbar, E_N(\hbar)})=  \hbar^d \wcal_{\hbar} \delta(x - y)\\ 
&= \int_{\R^d} \delta(x - v/2, x + v/2) e^{\frac{i}{\hbar} \xi \cdot v } dv \equiv 1. 
\end{align*}
 On the other hand, the absolute sum is infinite. To see this, we use the identity \eqref{E:Wigner-sp} and classical asymptotics of associated Laguerre functions. The sums in \eqref{EMPDEF} are weighted by $W_{\hbar, E_N(\hbar)}(x, \xi)$. We now show that the measures have infinite mass even if   we  $L^2$-normalize the Laguerre functions 
to form an orthonormal basis 
\begin{equation} \psi_N^{\alpha} = \left(\frac{2^{-\alpha} \Gamma(N +1)}{\Gamma(N + \alpha +1)}\right)^{\half} L_N^{\alpha}(\half r^2) e^{-\frac{1}{4} r^2}, \end{equation}
of  $L^2(\R_+, r^{2 \alpha +1} dr)$. By \eqref{E:Wigner-sp}, we want to show that
$$ \sum_{N = 1}^{\infty}   \left(\frac{2^{-\alpha} \Gamma(N +1)}{\Gamma(N + \alpha +1)}\right)^{\half}
   | e^{-  2H/\hbar}  L^{(d-1)}_N(4H/\hbar)| = \infty.$$
Since the exponential factor is independent of $N$, it may be removed from the sum. Moreover, the argument is independent of $N$ so it suffices to show that
 $$ \sum_{N =1}^{\infty}  \left(\frac{2^{-\alpha} \Gamma(N +1)}{\Gamma(N + \alpha +1)}\right)^{\half} |   L^{(d-1)}_N(x) | = \infty, \forall x > 0. $$ 
The relevant asymptotics are the Hilb asymptotics of the Bessel  regime, where $x$ is small relative to $N$,
 $$e^{-\frac{x}{2}} x^{\alpha/2} L_n^{(\alpha)}(x) = \frac{(n + \alpha)!}{n!} (N x)^{-\alpha/2} J_{\alpha}(2 \sqrt{Nx}) + \epsilon(x, n), $$
 where $$N = n + \frac{\alpha +1}{2}, \;\;\epsilon(x, n)  = \left\{\begin{array}{ll}
 x^{\alpha/2 + 2} O(n^{\alpha}), & 0 < x < \frac{c}{n}, \\ &\\
 x^{5/4} O(n^{\alpha/2 - 3/4}), & \frac{c}{n} < x < C. \end{array} \right. $$
Note that $\frac{\Gamma(n + \alpha +1)}{\Gamma(n)} \simeq n^{\alpha}$,
so $\frac{(n + \alpha)!}{n!} (N )^{-\alpha/2} \simeq n^{\alpha/2}.$
The $L^2$-normalizing constant puts in a factor asymptotic to $N^{-\alpha/2}$, cancelling the power in the Hilb asymptotics. In addition, we have the  Bessel asymptotics of fixed order and large argument. When $0 < x < \sqrt{\alpha +1}$,   $J_{\alpha}(x) \sim \frac{1}{\Gamma(\alpha +1)} (\frac{x}{2})^{\alpha}).$ When $x >> |\alpha^2 - \frac{1}{4}|$ then $J_{\alpha}(x) \sim \sqrt{\frac{2}{\pi x}} \cos(x - \frac{\alpha \pi}{2} - \frac{\pi}{4}). $ Clearly, the latter  is the relevant form,   and it  puts in an additional factor of $N^{-\half}$.
It follows that the summands  are of asymptotic order  $N^{-\half},$ and therefore the mass is infinite. A fortiori the mass of the measures \eqref{EMPDEF} is infinite.

To complete the proof of Proposition \ref{INFINITEPROP} we need to show
that the mass of \eqref{EMPDEF} is finite on one-sided infinite intervals $[-\infty, \tau]$. 
This is due to  the `two-sidedness' of the Airy function. The main point (see Section \ref{HEURISTICSECT}) is that terms with $E_N(\hbar) < H(x, \xi)$ are exponentially small and may be neglected. Therefore, 
$$\left| \mu_h^{(x, \xi)} \right| [-\infty, \tau] = \sum_{N: E_N(\hbar) \in [-\infty, \tau]}
|W_{\hbar, E_N(\hbar)}(x, \xi)| \simeq  \sum_{N: H(x, \xi) \leq E_N(\hbar) \leq \tau}
|W_{\hbar, E_N(\hbar)}(x, \xi) | < \infty$$
 for any $\hbar > 0$. 


\subsection{Wigner distributions of the propagator}
 
 Taking  \eqref{E:WignerK} to be  the propagator $U_\hbar(t)$ of \eqref{Hh},  its Wigner transform \eqref{WIGU} may be calculated explicity by using the Mehler formula for $U_\hbar(t,x,y)$ in the position representation (see e.g. \cite{F})
\begin{equation}  \label{E:Mehler}
 U_h(t, x,y) =  e^{-\ihbar t H_\hbar}(x,y)=  \frac{1}{(2\pi i \hbar \sin t)^{d/2}}
 \exp\left[ \frac{i}{\hbar}\left(
 \frac{\abs{x}^2 + \abs{y}^2}{2} \frac{\cos t}{\sin t} - \frac{x\cdot
 y}{\sin t} \right) \right],
\end{equation}
where $t \in \R$ and $x,y \in \R^d$. The right hand side is singular at $t=0.$ It is well-defined as a distribution, however, with $t$ understood as $t-i0$. Indeed, since $\widehat{H}_\hbar$ has a positive spectrum the propagator
$U_\hbar$ is holomorphic in the lower half-plane and $U_\hbar(t, x, y)$ is the boundary value of a holomorphic function in $\{\Im t < 0\}$.

\section{\label{WIGNERINTRO} Uniform asymptotics of Wigner functions of spectral projections: Proof of Theorem  \ref{ELEVELLOC}  }

In this section we prove Theorem \ref{ELEVELLOC} about asymptotics for the Wigner-Weyl sums
\[W_{\hbar, f, E}(x,\xi)=  \sum_N f(\hbar^{-1}(E-E_N(\hbar)) W_{\hbar, E_N(\hbar)}(x,\xi)\]
 defined in \eqref{UhfDEF}, where $f$ is a Schwartz function. Our starting point is to combine \eqref{E:smoothed-sum} with $\delta(\hbar)=\hbar$ with Proposition \ref{P:Wigner-prop}, which together read
\begin{equation}
W_{\hbar, f, E}(x,\xi)= \int_\R e^{-itE/\hbar} \widehat{f}(t) A_\hbar(t)\exp\lr{-\frac{i}{\hbar}\Psi(t,x,\xi)}\frac{dt}{2\pi}.\label{E:elevelloc-1}
\end{equation}
Here,
\[A_\hbar(t)=\lr{2\pi \hbar \cos(t/2)}^{-d},\qquad \Psi(t,x,\xi)=2H\tan(t/2),\quad H=H(x,\xi)=\frac{\abs{\xi}^2+\abs{\xi}^2}{2}.\]
Note that this distribution is actually a smooth function away from $t\in \set{\pi, 3\pi, \ldots}.$ By Lemma \ref{LOCLEM}, it remains to evaluate 
\begin{equation}
  \label{E:elevelloc-3}
  \int_\R e^{-itE/\hbar} \widehat{f}(t)\eta_\hbar(t) e^{-itE/\hbar}\ucal_{\hbar}(t,\rho)\frac{dt}{2\pi}~=~  \int_\R  \widehat{f}(t)\eta_\hbar(t) A(t)e^{-\frac{i}{\hbar}\lr{\Psi(t,E,\rho)-tE}}\frac{dt}{2\pi},
\end{equation}
where $\eta_\hbar:\R\gives [0,1]$ is a smooth, periodic cut-off function with
\[\eta_\hbar(t)=
\begin{cases}
  1,&\quad \abs{t-(2k+1)\pi}>2\hbar^\delta,\,\,\forall k\in \Z\\
  0,&\quad \abs{t-(2k+1)\pi}<\hbar^\delta,\,\,\text{for some }k\in \Z\\
\end{cases}.
\]
We do this by stationary phase. The critical point equation for $\Psi(t, E, \rho)-tE$ as a function of $t$ is
\begin{equation} \label{CPE1} 
H_E=\cos^2\lr{\frac{t}{2}},\qquad H_E:=\frac{\rho}{2E}.
\end{equation}
If $H_E>1$ (i.e. the point $(x,\xi)$ at which we evaluate \eqref{E:elevelloc-1} is outside the energy surface), there are no real critical points for $\Psi$ and the lemma of non-stationary phase shows that \eqref{E:elevelloc-3} is $O(\hbar^\infty)$ as desired. If $H_E\in (0,1),$ we use the method of stationary phase (Lemma \ref{L:SP LO}). The critical points of the phase $\Psi(t,E,\rho)-tE$ in \eqref{E:elevelloc-3} are
\begin{align*}
t_{+,\pm,k}&= 4\pi k \pm 2\cos^{-1}\lr{H_E^{1/2}},\quad t_{-,\pm,k}= 4\pi \lr{k+1/2} \pm 2\cos^{-1}\lr{H_E^{1/2}},\qquad k \in \Z.
\end{align*}
where $\cos^{-1}:[-1,1]\gives [0,\pi]$ and
\[\cos(t_{\pm_1,\pm_2, k})=\pm_1 H_E^{1/2},\qquad \sin(t_{\pm_1,\pm_2, k})=\pm_1\pm_2 \lr{1-H_E}^{1/2}.\]
 Notice that since $H_E>0,$ all the critical points are a bounded distance away from $\set{(2k+1)\pi,\, k\in \Z}$ and hence are inside the support of $\eta_\hbar$ for all $\hbar$ sufficiently small. The stationary phase analysis of \eqref{E:elevelloc-3} around $t_{\pm,\pm, k}$ is the same for all $k,$ so let us focus on $k=0.$ Since
\[\rho = 4E H_E,\]
we have for every $k$ that
\[\Psi(t_{\pm_1,\pm_2,k})=4EH_E\frac{\pm_1\pm_2\lr{1-H_E}^{1/2}}{\pm_1H_E^{1/2}}= \pm_24E\lr{H_E(1-H_E)}^{1/2}\]
and
\[\Psi''(t_{\pm_1,\pm_2, k})=\frac{\rho}{2}\frac{\sin(t_{\pm_1,\pm_2,k}/2)}{\cos^3(t_{\pm_1,\pm_2,k}/2)}=\pm_2 2E\lr{H_E^{-1}-1}^{1/2}\]
Note that $(2\pi \hbar)^{d-1/2}$ times the intergral in \eqref{E:elevelloc-3} becomes
\begin{align*}
 \lr{2\pi \hbar}^{-1/2} \int_\R \frac{\widehat{f}(t)\eta_\hbar(t)}{\cos(t/2)^d} e^{\frac{i}{\hbar}\lr{-\Psi(t,E,\rho)+tE}}\frac{dt}{2\pi}.
\end{align*}
Applying the stationary phase expansion (Lemma \ref{L:SP LO}), completes the proof (we obtain an error $\hbar^{1-2\delta}$ from the fact that after expanding to order $k$ in stationary phase, the remainder is a universal constant times the supremum of the $\lr{k+1}^{st}$ derivative of the amplitude, and each derivative of $\eta_\delta$ give a power of $\hbar^{-\delta}.$).  \hfill $\square$

\section{Interface asymptotics for smooth $\hbar^{2/3}$-localized Weyl sums:  Proof of Theorem \ref{RSCOR} }\label{RSCOR-pf}

We now prove Theorem \ref{RSCOR}. Our starting point is \eqref{E:smoothed-sum} with $\delta(\hbar)=\hbar:$
\[W_{f,\hbar, 2/3}(x,\xi)=\sum_{N\geq 0} f\lr{\hbar^{-2/3}\lr{E-E_N(\hbar)}}W_{\hbar, E_N(\hbar)}(x,\xi)=\int_{\R} \widehat{f}(t) e^{\frac{i}{\hbar^{2/3}}tE} \ucal(t\hbar^{1/3},x,\xi)\frac{dt}{2\pi}.\]
Taylor expanding the result of Proposition \ref{P:Wigner-prop} we obtain the following pointwise asymptotics
\begin{equation} \label{u13}
 e^{\frac{i}{\hbar^{2/3}}tE}\ucal(t\hbar^{1/3},x,\xi) = (2\pi \hbar)^{-d}e^{i\left[t\frac{E-H}{\hbar^{2/3}}-\frac{t^3}{12}H\right]}\lr{1+O((1+\abs{t})^5\hbar^{2/3})},
\end{equation}
and, moreover, that the left hand side has a complete asymptotic expansion in ascending powers of $\hbar^{2/3}.$ Since $\widehat{f}$ is Schwartz, touine Taylor expansion shows that $W_{f,\hbar, 2/3}(x,\xi)$ has a complete asymptotic expansion in ascending powers of $\hbar^{2/3}$ as well. To compute the leading term, recall that 
\begin{equation*} H=H(x,\xi)=E + u \lr{\frac{\hbar}{2E}}^{2/3}.\end{equation*} 
This assumption and \eqref{u13} imply that
\begin{equation} \label{WfFT} 
W_{f,\hbar, 2/3}(x,\xi)=(2\pi \hbar)^{-d}\int_{\R} \widehat{f}(t)e^{-i\left[ \frac{tu}{(2E)^{2/3}} + \frac{t^3}{12}E\right]}\frac{dt}{2\pi} \lr{1+O((1+\abs{u})\hbar^{2/3})},\end{equation}
Changing variables $t\mapsto -C_E^{-1} t$, with $C_E= (4/E)^{1/3}$, we arrive at the representation
\[W_{f,\hbar, 2/3}(x,\xi)=-(2\pi \hbar)^{-d}\int_{\R} C_E\widehat{f}(-t)(C_E t)e^{i\left[ \frac{tu}{E} + \frac{t^3}{3}\right]}\frac{dt}{2\pi}\lr{1+O((1+\abs{u})\hbar^{2/3})}.\]
Using that that inverse Fourier transform of $-C_E\widehat{f}(-tC_E)$ in the $t\gives \lambda$ variables evaluated at $\lambda$ is $f(-\lambda/C_E)$, that the Airy function is the inverse Fourier transform of $e^{it^3/3}$, and that the Fourier transform is an isometry on $L^2$, we therefore find
\[W_{f,\hbar, 2/3}(x,\xi)=(2\pi \hbar)^{-d}\int_{\R} f( -\lambda/C_E ) \Ai\lr{\lambda  + \frac{u}{E}} d\lambda+O((1+\abs{u})\hbar^{-d+2/3}),\qquad C_E=(4/E)^{1/3},\]
completing the proof of Theorem \ref{RSCOR}.

\section{Interface asymptotics for  sharp $\hbar^{2/3}$-localized Weyl sums: Proof of Theorem \ref{SHARPh23INTER}  \label{2/3sumSect}}
In this section we consider the sharp $\hbar^{2/3}$ localized Weyl sums \eqref{W23EDEF} around an energy level $E$ and consider the interface scaling around the energy surface \eqref{SIGMAEDEF}. Recall that
\[W_{\hbar, E_N(\hbar)}(x,\xi)=\int_{S^1}e^{it(E_N(\hbar)-E)/\hbar} e^{itE/\hbar}\ucal_{\hbar}(t,x,\xi)\frac{dt}{2\pi}.\]
We claim that in the regime
\begin{equation}
H(x,\xi) = E + u\lr{\frac{\hbar}{2E}}^{2/3},\qquad E_N(\hbar)-E=O(\hbar^{1-\epsilon})\label{E:surface-scaling}
\end{equation}
we also have for any $\delta>0$
\[
W_{\hbar, E_N(\hbar)}(x,\xi)=\int_{S^1}\chi(\hbar^{-2/3+\delta}t)e^{it(E_N(\hbar)-E)/\hbar} e^{itE/\hbar}\ucal_{\hbar}(t,x,\xi)\frac{dt}{2\pi}+O(\hbar^\infty)
\]
where $\chi$ is a smooth cutoff the is identically $1$ near $0$ and supported in $[-1,1].$ To see this, we begin with the exact distributional formula
\[
W_{\hbar, E_N(\hbar)}(x,\xi)=\int_{S^1}e^{itE_N(\hbar)/\hbar}\ucal_{\hbar}(t,x,\xi)\frac{dt}{2\pi}=\int_{S^1}e^{it(E_N(\hbar)-E)/\hbar}e^{itE/\hbar}\ucal_{\hbar}(t,x,\xi)\frac{dt}{2\pi}.
\]
As in Lemma \ref{LOCLEM}, we can localize away from the singular $t=\pi$ for the amplitude of $\ucal_\hbar(t,x,\xi)$ at the cost of an error of size $O(\hbar^\infty).$ Then, on the support of $1-\chi(\hbar^{-2/3+\delta}t)$ and away from $t=\pi$, there are no critical points for the combined phase $itE-i\tan(t/2)\rho$, and hence, using the assumption from \eqref{E:surface-scaling} that $E-E_N(\hbar)=O(\hbar^{1-\epsilon})$ we may apply the lemma of non-stationary phase. Thus, to prove Theorem \ref{SHARPh23INTER}, it is enough to prove if for the localized integral
\begin{equation}
W_{2/3,E,\lambda_{\pm}}^{\chi}(x,\xi):=\int_{\R} \chi(\hbar^{-2/3+\delta}t)\frac{e^{itn_+}-e^{itn_-}}{e^{it}-1} e^{itE/\hbar} \ucal_\hbar(t,x,\xi)\frac{dt}{2\pi}+O(\hbar^{\infty}),\label{E:loc-interface}
\end{equation}
where the notation $n_{\pm}$ is from the definition \eqref{W23EDEF} of the localized Wigner function and we have $n_+-n_- \simeq \hbar^{-1/3}.$ Using that for any $a,b$
\[e^{ita}-e^{itb}=it\int_a^b e^{it\lambda}d\lambda,\]
rescaling $t\mapsto t\hbar^{1/3}$, and Taylor expanding the phase and amplitude in $e^{itE/\hbar}\ucal_\hbar(t,x,\xi)$, we find that that $W_{2/3,E,\lambda_{\pm}}^{\chi}(x,\xi)$ equals
\[(2\pi\hbar)^{-d}\int_{\R}{\bf 1}_{\set{s\in [\lambda_-,\lambda_+]}} \int_{\R} \chi(\hbar^{\delta}t)  e^{-it (\hbar^{2/3}\lr{E-H}-\lambda) + i (t^3/3)(E/4)}\frac{dtd\lambda}{2\pi}+O(\hbar^{-d+1/3-6\delta}).\]
Using that $H=H(x,\xi)=E+u\lr{\frac{\hbar}{2E}}^{2/3}$ and changing variables 
\[(\lambda,t)\mapsto (-\lambda,-tC_E^{-1})\qquad C_E=\lr{4/E}^{1/3}\]
shows that $W_{2/3,E,\lambda_{\pm}}^{\chi}(x,\xi)$ equals
\[(2\pi\hbar)^{-d}C_E\int_{\R}{\bf 1}_{\set{\lambda\in [-\lambda_+,-\lambda_-]}} \int_{\R} \chi(\hbar^{\delta}C_Et)  e^{i\lr{t (\lambda C_E+u/E) + t^3/3}}\frac{dtd\lambda}{2\pi}+O(\hbar^{-d+1/3-6\delta}+(1+\abs{u})\hbar^{-d+2/3-6\delta}).\]
Since (see \cite[Lem. 3.1]{HZZ16}) for any $\delta>0$
\[\int_{\abs{t}<\hbar^{-\delta}}e^{itu+it^3/3}\frac{dt}{2\pi}=\Ai(u)+O(\hbar^\infty),\]
we find because $\chi(\hbar^{\delta}t)=1$ for all $\abs{t}<\hbar^{-\delta},$ 
\[W_{2/3,E,\lambda_{\pm}}^{\chi}(x,\xi)=(2\pi\hbar)^{-d}C_E\int_{-\lambda_+}^{-\lambda_-}\Ai\lr{\frac{u}{E}+C_E\lambda} d\lambda + O\lr{\hbar^{-d+1/3-6\delta}+(1+\abs{u})\hbar^{-d+2/3-6\delta}}.\]
This completes the proof.

\section{Bulk asymptotics: Proofs of Propositions  \ref{smoothedbulk}
  and  \ref{pp:PBK-leading}}

   \subsection{Asymptotics for $\hat{f} \in C_0(\R)$: Proof of Proposition \ref{smoothedbulk}}\label{BULKfSECT}
  Fix $f\in \mathcal S(\R)$ with $\hat{f}\in C_0(\R).$ Relation \eqref{E:smoothed-sum} with $\delta(\hbar)=1$ and $E=0$ reads
\[[W_{\hbar, f}(x,\xi)=\sum_{N\geq 0} f(E_N(\hbar)) W_{\hbar, E_N(\hbar)}(x,\xi)=\int_{\R}\widehat{f}(t)\ucal_{\hbar}(-t\hbar,x,\xi) \frac{dt}{2\pi},\]
where we recall that $\ucal_{\hbar}$ is the Wigner function of the propagator. Proposition \ref{P:Wigner-prop} therefore yields
\[W_{\hbar, f}(x,\xi)= (2\pi \hbar)^{-d}\int_{\R} \widehat{f}(t)\lr{\cos(t\hbar/2)+i0}^{-d}e^{-\frac{2iH}{\hbar}\frac{\sin(-t\hbar/2)}{\cos(-t\hbar/2)}}\frac{dt}{2\pi},\quad H=H(x,\xi)=\frac{\norm{x}^2+\norm{\xi}^2}{2}.\]
For $\hbar$ sufficiently small, $\ucal_{\hbar}(-t\hbar, x,\xi)$ has no singularities on the support of $\widehat{f}.$ Taylor expanding in $\hbar$ therefore yields a complete asymptotic expansion for $W_{\hbar, f},$ as claimed. To complete the proof of Proposition \ref{smoothedbulk}, note that 
\[\ucal_\hbar(-t\hbar, x,\xi)=(2\pi \hbar)^{-d}e^{itH}\lr{1+O(\hbar^2)}.\]
Hence, the leading term in the expansion is 
\[(2\pi\hbar)^{-d}\int_{\R}\hat{f}(t) e^{itH}\frac{dt}{2\pi} =(2\pi\hbar)^{-d}f(H(x,\xi)),\] 
as desired.  

\begin{rem}
As discussed above, if $|D^{\ell} \widehat{f}| \in L^1(\R)$ for all $\ell \leq d$, then after integrating by parts $d$ times the integral becomes absolutely convergent. We may then calculate the limit $\hbar \to 0$ by dominanted convergence. The limit is $$H^{-d} \int_{\R} D^d \widehat{f}(t) e^{i tH} \frac{dt}{2\pi}= \int_{\R} \widehat{f}(t) e^{itH}\frac{dt}{2\pi} = f(H(x,\xi)), $$ as in the Schwartz case. \end{rem}

\subsection{Proof of Proposition  \ref{pp:PBK-leading} (i)}\label{PROOFBULK1} 
Proposition \ref{pp:PBK-leading} is formally Proposition \ref{smoothedbulk} in the case  $f = {\bf 1}_{[E_1, E_2]} (E_N(\hbar))$. 
It is sufficient to study the intervals $[0, E]$ (where we assume $H(x, \xi) > 0$), and so we assume henceforth that the interval has this form. Writing $f(E_N(\hbar)) := {\bf 1}_{[0, E]}(E_N(\hbar)),$ we are interested in
\[ W_{\hbar, [0,E]}(x,\xi)=\sum_{N: E_N(\hbar) \leq E} W_{\hbar, E_N(\hbar)}(x, \xi).\]
The endpoint of the sum is the largest  value $N(\hbar, E) \in {\mathbb N}$ for which
$E_N(\hbar)=\hbar(N + \frac{d}{2}) \leq E$, i.e. for which $N \leq \hbar^{-1} E - \frac{d}{2}$. Lemma \ref{NICEFORM} with $N_1=0$ and $N_2=N(\hbar, E)$ reads
\begin{equation} \label{EXACT-new}W_{\hbar, [0,E]}= \int_{S^1} \ucal_{\hbar}(t, x, \xi)  e^{itd/2} D_{\hbar, E}(t) \frac{dt}{2\pi},  \qquad D_{\hbar,E}(t):=\frac{e^{ i N(\hbar, E)t} -1}{e^{it} -1},\end{equation} 
where $D_{\hbar, E}$ is the Dirichlet kernel. Before turing to the detailed analysis of the expression in the previous line, note that three intervals in $t$ are important to this oscillatory integral expression for the pointwise Weyl sum $W_{\hbar, [0,E]}(x,\xi)$ as $\hbar\gives 0.$ The first contribution comes from the large peak of the Dirichlet kernel around $t=0$:
\[\abs{t}\leq N(\hbar,E)^{-1}\simeq \hbar\quad \Rightarrow \quad D_{\hbar, E}(t)\simeq N(\hbar, E)\simeq \hbar^{-1}. \]
Using that the expression for $\ucal_\hbar$ has a prefactor of size $\hbar^{-d},$ we will see that this interval gives an $\hbar^{-d}(1+O(\hbar))$ contribution to $W_{\hbar, [0,E]}$ as $\hbar\gives 0$ when $H(x,\xi)<E$ and is $O(\hbar^\infty)$ when $H(x,\xi)>E.$ The second contribution comes from the stationary phase points of $\ucal_\hbar.$ As we will see, when $H(x,\xi)<E,$ these points are well-separated from both $t=0$ and $t=\pi.$ A simple stationary phase argument will show that the resulting contribution is of size $O(\hbar^{-d+1/2}).$ In contrast, if $H(x,\xi)>E$, then there are no stationary phase points and this integral will also contribute $O(\hbar^\infty).$ Finally, we appeal to the argument from Lemma \ref{LOCLEM}  to show that that the apparent singularity $t=\pi$ of the amplitude $\cos(t/2)^{-d}$ in the expression for $\ucal_\hbar$ in fact is negligible (i.e. has size $O(\hbar^\infty)$) in the semi-classical limit $\hbar\gives 0.$

We now turn to the details. Fix a parameter $M>0$, which we will later take to be $\hbar^{-\epsilon}$ for $\epsilon>0$ sufficiently small, and let $\chi:\R\gives [0,1]$ be a smooth bump function:
\[\chi(t):=
\begin{cases}
  1,\qquad t \leq 1 \\
0,\qquad t > 2
\end{cases}.
\]
We first compute the contribution to $W_{\hbar, [0,E]}(x,\xi)$ from the peak of the Dirichlet kernel. Specifically, let us check that with $\epsilon$ sufficiently small and positive for $M= \hbar^{-\epsilon},$ we have
\begin{equation}
\int_{-\pi}^\pi \chi((M\hbar)^{-1} t)\ucal_{\hbar}(t, x, \xi)  e^{itd/2} D_{\hbar, E}(t) \frac{dt}{2\pi} =
\begin{cases}
(2\pi\hbar)^{-d}\lr{1 + O(M\hbar)},&~~ H(x,\xi)<E\\
O(\hbar^\infty),&~~ H(x,\xi)>E  
\end{cases}.\label{E:Bulk-Wig-Dir}
\end{equation}
This will turn out the be the dominant contribution to $W_{\hbar, [0,E]}(x,\xi)$ in the case $H(x,\xi)<E.$ The integrand in \eqref{E:Bulk-Wig-Dir} is supported on $\abs{t}\leq 2 M\hbar$. In this regime, Taylor expansion gives 
\[\ucal_\hbar(t,x,\xi)=(2\pi\hbar)^{-d}e^{-\frac{i}{\hbar}tH}(1+O((M\hbar)^2)+O(M^3\hbar^2)),\qquad H = H(x,\xi) =\frac{\norm{x}^2+\norm{\xi}^2}{2}.\]
Next, note that for any $\beta\in \R$
\[e^{i\beta t}-1=i\beta t \int_0^1 e^{i\beta t (1-s)}ds.\]
Hence, on the support of the integrand in \eqref{E:Bulk-Wig-Dir}, 
\[e^{itd/2}D_{\hbar, E}(t)= Ne^{\frac{i}{\hbar}tE}\int_0^1 e^{-iNts}ds (1+O(M\hbar)),\qquad N = N(\hbar, E)=\frac{E}{\hbar}-\frac{d}{2}.\]
Thus, using that $\abs{D_{\hbar, E}(t)}=O(\hbar^{-1})$ on the support of $\chi((M\hbar)^{-1}t)$, which has size $O(M\hbar),$ the integral in \eqref{E:Bulk-Wig-Dir} equals 
\[(2\pi\hbar)^{-d}\int_{-\infty}^\infty \int_0^N e^{\frac{i}{\hbar}t\left[(E-H)-s\right]} \chi\lr{(M\hbar)^{-1}t}  \frac{dsdt}{2\pi}\]
plus an error of size $O(M^2\hbar^{-d+1})+O(M^3\hbar^{-d+2}).$ Changing variables $t\mapsto t (\hbar M)$ and performing the $dt$ integral, the expresion in the previous line becomes
\[(2\pi\hbar)^{-d}(M\hbar)\int_0^N \widecheck{\chi}\lr{M\hbar (E-H - s)} ds,\]
where $\widecheck{\chi}$ is the inverse Fourier transform. Finally, changing variables $s\mapsto \hbar M(E-H - s)$, this integral becomes
\begin{equation}
(2\pi\hbar)^{-d}\int_{M(E-H-N)}^{M(E-H)} \widecheck{\chi}\lr{s} ds=(2\pi \hbar)^d\lr{\chi(0) + O(\hbar^\infty)}=(2\pi \hbar)^d\lr{1 + O(\hbar^\infty)},\label{E:inverse-fourier}
\end{equation}
provided $M=\hbar^\epsilon$ for any $\epsilon>0$ and assuming that $H<E$. In contrast, when $H>E,$ this integral is $O(\hbar^\infty)$ since $\widecheck{\chi}$ is rapidly decaying at infinity. This proves \eqref{E:Bulk-Wig-Dir}. The same argument as in Lemma \ref{LOCLEM} combined with \eqref{E:Bulk-Wig-Dir} shows that
\[W_{\hbar, [0,E]}= (2\pi \hbar)^{-d}(1+O(\hbar^{1-\epsilon}))+
\int_{-\pi}^\pi (1-\chi((M\hbar)^{-1} t))\psi_\delta(t)\ucal_{\hbar}(t, x, \xi)  e^{itd/2} D_{\hbar, E}(t) \frac{dt}{2\pi}
+O(\hbar^\infty),\]
where $\psi_\delta$ is a smooth cut-off function that localizes the integral away from a $\hbar^{\delta}$ neighborhood of $t=\pi.$ Thus, to complete the proof of Proposition \ref{pp:PBK-leading}, it remains to check
\begin{equation}
\int_{-\pi}^\pi (1-\chi((M\hbar)^{-1} t))\psi_\delta(t)\ucal_{\hbar}(t, x, \xi)  e^{itd/2} D_{\hbar, E}(t) \frac{dt}{2\pi} =O(\hbar^{-d+1/2})\label{E:Bulk-Wig-SP}
\end{equation}
where $M=\hbar^{\epsilon}$ and $\epsilon, \delta$ are sufficiently small. Our argument uses the method of stationary phase. We write
\[\ucal_{\hbar}(t, x, \xi)  e^{itd/2} D_{\hbar, E}(t)= \frac{e^{-\frac{i}{\hbar}(\Psi(t,x,\xi)-tE)}-e^{-\frac{i}{\hbar}\Psi(t,x,\xi)}}{(\cos(t/2)+i0)^{d}(e^{it}-1)},\quad \Psi(t,x,\xi)=\tan(t/2) (\norm{x}^2+\norm{\xi}^2).\]
On the support of $(1-\chi((M\hbar)^{-1}t))\psi(t)$, the phase function $\Psi$ has no critical points, while the amplitude has its $k^{th}$ derivative bounded by $\hbar^{k\max\set{\delta, \epsilon}}.$ Thus, the integral in \eqref{E:Bulk-Wig-SP} corresponding to term with $e^{-\frac{i}{\hbar}\Psi(t,x,\xi)}$ gives contribution $O(\hbar^\infty).$ In contrast, the critical point equation for $\Psi(t,x,\xi)-tE$ is
\begin{equation}
\cos^2(t/2)=H(x,\xi)/E.\label{E:crits-again}
\end{equation}
This has two real solutions bounded away from $t=0,\pi$ when $0<H(x,\xi)<E$ and has no real solutions if $H(x,\xi)>E$. In the former case, these critical points are easily seen to be non-degenerate and the usual method of stationary phase show that the contribution to $W_{hbar,[0,E]}(x,\xi)$ is of order $\hbar^{-d+1/2}$ when $\epsilon,\delta$ are sufficiently close to $0.$ In the latter case, the integral is $O(\hbar^\infty).$ This completes the proof of Proposition \ref{pp:PBK-leading}.\hfill $\square$

	\section{Interface asymptotics for bulk Wigner distributions: Proof of Theorem \ref{BULKSCALINGCOR}}
	

Throughout this section we fix an energy level $E$ and consider the energy window or shell $[0, E]$. We are interested in the interface asymptotics behavior of bulk Wigner functions at points $(x, \xi)$ which satisfy
\begin{equation} \label{ASSUME}H(x, \xi)= \frac{\abs{x}^2+\abs{\xi}^2}{2} = E + u\lr{\frac{\hbar}{2E}}^{2/3}.\end{equation}
Combining Lemma \ref{NICEFORM} gives
\[W_{\hbar, [0,E]}(x,\xi)=\int_{-\pi}^\pi \chi(t) D_{\hbar, E}(t)e^{itd/2} \ucal_\hbar(t,x,\xi)\frac{dt}{2\pi}+O(\hbar^\infty)\]
where 
\[D_{\hbar, E}(t)=\frac{e^{itN_E}-1}{e^{it}-1},\qquad N_E=E/\hbar - d/2\]
and $\chi$ is a smooth cutoff that is identically $0$ near $t=\pm \pi.$ Proposition \eqref{P:Wigner-prop} therefore shows that, up to an error of size $O(\hbar^\infty),$
\[W_{\hbar, [0,E]}(x,\xi)=(2\pi \hbar)^{-d}N_E \int_{-\pi}^\pi \chi_1(t)D_{\hbar, E}(t)e^{itd/2}\cos(t/2)^{-d} e^{-\frac{2i H}{\hbar}\tan(t/2)}\frac{dt}{2\pi},\]
where as before $H=H(x,\xi).$ Note that both phase functions $2H\tan(t/2)$ and $2H\tan(t/2)-tE$ have no critical points outside $\abs{t}\leq C\hbar^{1/3}\abs{u}^{1/2},$ where $C$ is a fixed sufficiently large constant. Thus, we may further localize:
\[W_{\hbar, [0,E]}(x,\xi)=(2\pi \hbar)^{-d}\int_{-\pi}^\pi \chi(t\hbar^{-1/3+\epsilon}\abs{u}^{-1/2})D_{\hbar, E}(t)e^{itd/2}\cos(t/2)^{-d} e^{-\frac{2i H}{\hbar}\tan(t/2)}\frac{dt}{2\pi}+O(\hbar^\infty)\]
As above we use that
\[D_{\hbar, E}(t)=-\frac{it}{e^{it}-1} N_E\int_0^1 e^{itN_E(1-s)}ds=-(1+O(t))e^{itE/\hbar}\int_0^{N_E} e^{-its}ds\]
to find that, up to errors of size $O(\hbar^\infty)$ the normalized Wigner function $(2\pi\hbar)^d W_{\hbar, E}(x,\xi)$ equals
\[N_E\int_0^{1}\int_{-\pi}^\pi \chi(t\hbar^{-1/3}\abs{u}^{-1/2})(1+O(t))e^{\frac{i }{\hbar}\lr{2H\tan(t/2)+t(s-E)}}\frac{dt}{2\pi}ds.\]
Taylor expanding yields 
\[2H\tan(t/2)+t(s-E)=t(H-E+s) ~+~Ht^3/12 + O(t^5).\]
Thus, $(2\pi \hbar)^d W_{\hbar, [0,E]}(x,\xi)$ is
\[N_E\int_0^1 \int_{-\infty}^\infty \chi(t\hbar^{-1/3+\epsilon}\abs{u}^{-1/2})e^{\frac{i}{\hbar}\lr{t(H-E+s)+Et^3/12}}\lr{1+R(t,s)}\frac{dt}{2\pi}ds,\]
where the remainder $R$ satisfies
\[R(t,s)=O(t)+O(t^5/\hbar)+O(t^3\hbar^{-1/3}\abs{u}).\]
Since the integrand vanishes when $\abs{t}\geq \hbar^{1/3-\epsilon}\abs{u}^{1/2},$ we find that $(2\pi \hbar)^d W_{\hbar, [0,E]}(x,\xi)$ is
\[N_E\left[\int_0^1 \int_{-\infty}^\infty \chi(t\hbar^{-1/3+\epsilon}\abs{u}^{-1/2})e^{\frac{i}{\hbar}\lr{t(H-E+s)+Et^3/12}}\frac{dt}{2\pi}ds+O(\hbar^{1/3}\abs{u}^{1/2})+O(\hbar^{2/3}\abs{u}^{5/2})\right].\]
Changing variables to 
\[T=(4\hbar/E)^{1/3}t,\qquad\sigma = s(\hbar/2E)^{2/3}\]
and using that $\hbar N_E = E + O(\hbar),$ we obtain the following expression for $(2\pi \hbar)^d W_{\hbar, [0,E]}(x,\xi):$
\[\int_{u/E}^{u/E+(2E/\hbar)^{2/3}} \int_{-\infty}^\infty \chi(T\hbar^{\epsilon}\abs{u}^{-1/2})e^{\frac{i}{\hbar}\lr{T\sigma+T^3/3}}\frac{dt}{2\pi}d\sigma= \int_0^{\infty}\Ai(u/E+t)dt\]
plus an error of size $ O(\hbar^{1/3}\abs{u}^{1/2})+O(\hbar^{2/3}\abs{u}^{5/2}).$ This completes the proof of Theorem  \ref{BULKSCALINGCOR}.

\section{\label{APPENDIX} Appendix}

\subsection{Stationary Phase Expansion}\label{SO} We recall here the following simple version of the  stationary expansion, which we use in several proofs.
\begin{Lem}[\cite{Hor} Theorem 7.7.5]\label{L:SP LO}
Suppose $a,S\in \mathcal S(\R)$ and $S$ is a complex-valued phase function such that $\Im
S|_{\supp(a)} \geq 0$ with a unique non-degenerate critical point at $t_0\in \supp(a)$
satisfying $\Im S(t_0)=0.$ Define
\[I(h)=(2\pi h)^{-1/2}\int_{\R}e^{iS(x)/h}a(x)dx.\]
Then
\begin{align}
\label{E:Stationary Phase 2}  I(h) = C(S) \left[a(t_0)+O(h)\right],\qquad 
C (S)= e^{i \frac{\pi}{4} \text{sgn} S''(0)} \lr{\frac{2\pi
h}{\abs{S''(t_0)}}}^{1/2}.
\end{align}
\end{Lem}

\subsection{\label{AIRYAPP} Appendix on the Airy function}
The Airy function is defined by,

$$Ai(z) = \frac{1}{2 \pi i} \int_L e^{v^3/3 - z v} dv, $$
where $L$ is any contour that beings at a point at infinity in the sector $- \pi/2 \leq \arg (v) \leq - \pi/6$ and ends
at infinity in the sector $\pi/6 \leq \arg(v) \leq \pi/2$.  In the region $|\arg z| \leq (1 - \delta) \pi$
in $\C - \{\R_-\}$ write $v = z^{\half} + i t ^{\half}$ on the upper half of L and $v = z^{\half} - i t^{\half}$
in the lower half. Then
\begin{equation} \label{AIRYASYM} \Ai(z) = \Psi(z) e^{- \frac{2}{3} z^{3/2}}, \;
\rm{
with}\;
\Psi(z) \sim z^{-1/4} \sum_{j = 0}^{\infty} a_j z^{- 3j/2}, \;\; a_0 = \frac{1}{4} \pi^{-3/2}. \end{equation}

\end{document}